\pdfoutput=1
\documentclass{article}

\usepackage[english]{babel}
\usepackage[letterpaper,top=2cm,bottom=2cm,left=3cm,right=3cm,marginparwidth=1.75cm]{geometry}

\usepackage[hidelinks]{hyperref}
\usepackage{varioref}

\usepackage{csquotes}

\usepackage[style=apa,backend=biber,natbib=true]{biblatex}
\usepackage{fancyref}
\usepackage{amsmath}
\usepackage{graphicx}
\usepackage{float}
\usepackage{array,threeparttable,longtable,booktabs,makecell}

\usepackage[capitalise,nameinlink,noabbrev]{cleveref}
\newcommand{\footremember}[2]{%
    \footnote{#2}
    \newcounter{#1}
    \setcounter{#1}{\value{footnote}}%
}
\newcommand{\footrecall}[1]{%
    \footnotemark[\value{#1}]%
} 

\title{Beyond Citations: Measuring Idea-level Knowledge Diffusion from Research to Journalism and Policy-making}
\author{%
  Yangliu Fan\footremember{wi}{Weizenbaum Institute, Berlin, Germany}\footremember{fu}{Institute for Media and Communication Studies, Freie Universität Berlin, Berlin, Germany}
  \and Kilian Buehling\footrecall{wi} \footrecall{fu}
  \and Volker Stocker\footrecall{wi}\footnote{Technische Universität Berlin, Berlin, Germany}
  \footnote{Corresponding author: Yangliu Fan, yangliu.fan@weizenbaum-institut.de}
}

\date{}

\begin{document}
\typeout{JOBNAME=\jobname  EXPECTED=main}
\maketitle

\begin{abstract}
Despite the importance of social science knowledge for various stakeholders, measuring its diffusion into different domains remains a challenge. This study uses a novel text-based approach to measure the \textit{idea-level} diffusion of social science knowledge from the research domain to the journalism and policy-making domains. By doing so, we expand the detection of knowledge diffusion beyond the measurements of direct references. Our study focuses on media effects theories as key research ideas in the field of communication science. Using 72,703 documents (2000-2019) from three domains (i.e., research, journalism, and policy-making) that mention these ideas, we count the mentions of these ideas in each domain, estimate their domain-specific contexts, and track and compare differences across domains and over time. Overall, we find that diffusion patterns and dynamics vary considerably between ideas, with some ideas diffusing between other domains, while others do not. Based on the embedding regression approach, we compare contextualized meanings across domains and find that the distances between research and policy are typically larger than between research and journalism. We also find that ideas largely shift roles across domains---from being the theories themselves in research to sense-making in news to applied, administrative use in policy. Over time, we observe semantic convergence mainly for ideas that are practically oriented. Our results characterize the cross-domain diffusion patterns and dynamics of social science knowledge at the idea level, and we discuss the implications for measuring knowledge diffusion beyond citations. 
\end{abstract} \hspace{10pt}
Keywords: knowledge diffusion, knowledge transfer, communication science, journalism, policy, word embeddings

\section{Introduction}

The transfer of knowledge from academia to non-academic stakeholders and practitioners is critical for societal progress (\cite{Weiss1979, Cohen2003, David2007, Zawdie2010}). In recent years, knowledge transfer has been described as the \textit{Third Mission} of research institutions---alongside research and teaching---serving to strengthen their roles in innovation and regional and global development processes (\cite{Zawdie2010}). While researchers have adopted various strategies to communicate or co-produce knowledge, such as patents, licensing, formal and informal research collaborations, and meetings or consulting (\cite{Cohen2003, David2007}), transfer channels and outputs vary significantly across research fields and disciplines. Effective knowledge transfer and diffusion, therefore, require context-specific strategies. For instance, in STEM (science, technology, engineering, and mathematics) fields, patents, technology licenses, or spin-offs are common transfer methods (\cite{David2007, Yin2022}). Social science researchers rely on different channels and outputs, such as magazines and newspapers, multi-stakeholder events or forums, consultations, or novel approaches to science communications (e.g., via blogs or social media) (\cite{Weiss1979, Hallett2019, Yin2022, Cao2025}). 

Importantly, insights from social science research are often less tangible, and measuring such knowledge diffusion is inherently challenging. It is therefore unsurprising that empirical investigations remain scant. Moreover, most existing research (e.g., \cite{Yin2022}) is limited to citation patterns, focusing on direct references to a specific publication. Such citation patterns, however, indicate \textit{whether} research is cited, but not \textit{how} it is used and \textit{what} parts of its meaning are retained. 
Diffusion channels and paths of social science knowledge and understandings are complex and may be circuitous. That is, they cannot always be captured as a direct citation or an instrumental use. Rather, diffusion may manifest as a conceptual frame, interpretant, or meta-discourse that (re-)shapes frames, ideas, and orientations through which actors perceive societal challenges in the longer term (\cite{Weiss1979, Weiss1980, Daviter2015, Hallett2019}). Drawing on the \textit{enlightenment} model of the utilization of social science research (\cite{Weiss1979}) and the concept of \textit{public idea} (\cite{Hallett2019}), we conceptualize \textit{idea-level} diffusion as a process in which social science ideas (i.e., knowledge generated in academic domains, that is tangible through distinctive concepts and theories)  diffuses to other societal domains. That is, knowledge from social sciences permeates society over time through \textit{social science ideas} with ways of making sense of complex phenomena in society (See Section Theoretical background for details).  

In this paper, we develop a novel method to track such idea-level diffusion of social science knowledge into the journalism and policy-making domains. As a case study, we focus on media-effects theories, including theoretical thoughts and traditions that are central to communication science (\cite{Neuman2011}). Building on the 33 named concepts of media-effects research outlined in Neuman and Guggenheim’s six-stage model, we treat these concepts (e.g., \textit{agenda-setting} and \textit{public sphere}) as tractable units of social science ideas. Next, we use a text-based approach to analyze the adoption and diffusion patterns of these ideas into the journalism and policy domains. This enables us to detect cross-domain diffusion and assess \textit{meaning retention}---that is, when non-academic stakeholders use a concept, do they use it in ways semantically close to its scholarly baseline, or do they (re-)contextualize it within domain-specific frames? By doing so, we address a key gap: while existing scientometric studies (e.g., \cite{Haunschild2017,Yin2021,Yin2022,Liu2022,Bornmann2022,Ren2023,Cao2025}) have mapped citation flows into media and policy documents, we lack empirical research regarding how social science ideas diffuse into other domains, and whether their meaning is retained or rather unrelated between research and other domains.

More specifically, our method first counts the mentions of relevant ideas across the three domains, i.e., research, journalism, and policy-making. Next, we use the embedding regression approach (\cite{Rodriguez2023}) to analyze how the use of these ideas varies across different domains and over time. This approach measures whether tokens (the distinctive names of media effects theory concepts in our case) are used differently across sub-corpora and whether those differences are statistically significant. Additionally, we study \emph{(re-)contextualization} by identifying the nearest neighborhoods in the learned embedding space that co-occur with each concept in similar contexts in each domain (See Section Methodology for details). 

Analyzing a total of 72,703 documents (2000-2019) from research, journalism, and policy domains that mention these prominent ideas, we find that the idea-level diffusion is largely heterogeneous and complex, with only a subset of social science ideas diffusing to news and policy domains. Moreover, we find a nontrivial semantic shift outside academia. Specifically, we find that in the policy domain, the semantic meaning of media effects theory concepts usually diverges farther from the use in the research domain than does the use in the journalism domain. Notably, there are two exceptions to ideas that are codified in policy instruments and programs (practically-oriented). Further, we observe a general role shift of social science ideas---from being the theories themselves in the research domain to becoming a sense-making device used broadly in the news domain, to being further narrowed down in an applied, administrative use in the policy domain. We further characterize different types of ideas based on their diffusion patterns and dynamics: practically-oriented (e.g., \textit{social networks} and \textit{social capital}) and interpretive ideas (e.g., \textit{public sphere} and \textit{social identity}), as well as polysemes (e.g., \textit{persuasion} and \textit{priming}), which differ in the magnitude of their semantic shift as they diffuse across different domains. 

In sum, this paper studies how distinct social science ideas diffuse from the research domain to the journalism and policy domains. By focusing on the idea-level semantic distance, our approach complements existing citation- and mention-based measures with a view of diffusion that attends to meaning and context, especially when direct references are absent. Our study offers a novel measurement framework that can be potentially generalizable for future studies on idea-level knowledge diffusion in a wide range of different contexts and motivates further (meta-)theoretical explorations on how various societal domains interact through \textit{knowledge}.

\section{Theoretical background} 
In the following, we first review existing scientometric approaches to knowledge diffusion beyond academia. Next, we give a brief overview of the core concepts and theoretical insights that have informed our idea-level framework. In particular, the theoretical background of our framework draws on various domains of research, including sociology, political science, and communication science.  

\subsection{Existing scientometric approaches to cross-domain knowledge diffusion}

A growing number of scientometric studies have examined the broader impacts of scientific knowledge and how it diffuses outside the research domain. Essentially, these studies aim to find new ways to measure the \textit{impact} of research beyond academia, thereby addressing the limitation of traditional citation analyses measuring the impact exclusively within the research domain (\cite{Haunschild2017, Bornmann2022}). By expanding the focus on the broader impact of science, these studies explore how research influences real-world settings (\cite{Vilkins2017}), thus shedding the ivory tower image of research and research institutions (\cite{David2007}) and providing evidence to legitimize public funding decisions (\cite{Yin2022}). Research on the policy impact of research (e.g., through \textit{evidence-based policy making} (\cite{Black2001})) can, for example, shed light on how scientific findings have been used to address urgent societal challenges, with COVID-19 (\cite{Yin2021}) and climate change (\cite{Bornmann2016, Bornmann2022}) as two prominent examples. 

By tracking scholars' increased use of digital scientific communication channels, altmetrics (an umbrella term for alternative metrics) databases (e.g., Altmetric and Overton) have provided scientometricians with massive data on a global scale, such as views, downloads, social media, mainstream media, and policy-related mentions (\cite{Priem2012, Haunschild2017,Liu2022,Szomszor2022}). Using these data, researchers have measured the presence or absence of external impact---that is, whether and how often research papers are mentioned in external sources, e.g., news and policy documents. Specifically, existing research has examined the origins of the scientific research in terms of article types (e.g., review articles and research articles), journals, disciplines (or cross-disciplinarity), and the credibility of the sources (preprints or peer-reviewed journal articles) they first appeared in (\cite{Bornmann2022, Pinheiro2021, Ren2023, Yin2021, Bornmann2016, Vilkins2017}), the institutional conditions of the audiences, such as news and social media, think tanks, governments (including mentions in the government reports and legislative documents), inter-governmental organizations (IGOs) and other boundary organizations (\cite{Cao2025, Yin2021, Bornmann2022, Liu2022, Yin2022}), as well as the channels (such as the science-policy interface) and the scope and efficiency of such cross-domain knowledge diffusion (\cite{Ren2023, Yin2021}). Furthermore, research has also attempted to interpret different types of mentions of scholarly work (\cite{Yu2023}) and assess the accuracy of such mentions provided by these databases (\cite{Yu2022}). More recent work has also traced citation pathways to study the process of impact by distinguishing direct and indirect impact and identifying the presence or absence of reinforcement effects (\cite{Cao2025}). 

While existing citation- and mention-based approaches provide valuable insights and further approaches to understanding cross-domain knowledge diffusion, such approaches are inherently limited because they are merely a proxy of the actual use of the cited research results (\cite{Yin2021,Yu2023}). Mentions in policy documents, for instance, do not always prove that the cited research has influenced the policy process (\cite{Newson2018}), nor do they reflect the actual non-linear and complex pathways of research impact on different societal domains (\cite{Cao2025}). Moreover, previous research (\cite{Bornmann2022}) has found that some policy documents, particularly laws, do not contain any references. Therefore, citation- and mention-based indicators might not be able to detect relevant use patterns.   
In other words, these indicators may show whether research is mentioned (on the article level) but do not capture \emph{how} it is used, \emph{what} is retained, and \textit{whether} it is interpreted differently or even distorted during the diffusion process. 

Motivated by these gaps in our understanding, this paper introduces a novel measurement framework that allows us to measure the meanings and contexts of cross-domain knowledge diffusion. Specifically, we use a text-based approach to trace and measure the diffusion of \textit{social science} research ideas---named concepts---which are often overlooked in existing diffusion studies mainly focusing on biomedical research. By focusing on the named concepts (e.g., \textit{public sphere}) as the trackable units of diffusion, we can capture the \textit{uncited} uptake of social science knowledge, expanding the detectability of knowledge diffusion beyond direct references. Before turning to the empirical application, the next section will first establish the theoretical grounding of our framework, drawing also connections to the literature on sociology, political science, and communication science. 

\subsection{Towards an idea-level framework of cross-domain knowledge diffusion: social science ideas in journalism and policy-making}

There is substantial evidence that many non-academic stakeholders use social science research (\cite{Hallett2019, Yin2022, Ren2023,Vilkins2017}). Journalists and policymakers, in particular, have been found to explicitly or implicitly draw on knowledge derived from social research and analysis in their professional practices (\cite{Weiss1979, Weiss1980, Hallett2019, Yin2022, Daviter2015}). As a result, social policies, government programs, and public discourse are often informed and shaped by social science research and understanding. 

Importantly, however, existing research (\cite{Weiss1980, Daviter2015}) has pointed out that it is rare for policy-makers to use a single social science study as a direct input or in an instrumental fashion. Rather, knowledge is often selected, aggregated, and transformed by journalists as knowledge brokers (\cite{Gesualdo2020}) or mediators (\cite{Brggemann2014}). Furthermore, knowledge may \textit{creep} into policy deliberations in a diffuse and indirect manner, based on a substantial body of research results (\cite{Khazragui2015, Weiss1980}). During this process of \textit{knowledge creep}, social science research \textit{permeates} society over a longer period of time through generalizations and orientations that shape the way people define issues and perceive problems (\cite{Weiss1979}). Through journalism and other mediating institutions, social science ideas and orientations also enter public discourse and debates, where they guide informed publics and influence how people think about complex phenomena (\cite{Hallett2019, Gesualdo2020, Yarnall2017}). 

In the policy research literature, the \textit{enlightenment} model of social science knowledge use in policy-making, developed by Carol Weiss, is considered one of the dominant theoretical perspectives (\cite{Daviter2015}). Essentially, this perspective shifts scholarly attention from a narrow, instrumental, or concrete knowledge use (in a direct and measurable manner) to a broader, \textit{conceptual} use of social science knowledge during policy-making processes (\cite{Weiss1980}). Such conceptual use is slow, indirect, and cumulative, but it may affects thinking in the long term and contributes to the understanding of the nature of social problems. This provides a more realistic account of how social science research produces societal impact, which is more suited to the nature of social science research outputs (compared to outputs from STEM, where practical action and direct implementation are more common). Based on interviews of 155 government officials in the United States, Weiss found that the conceptual knowledge use is more common than the narrow, instrumental, or concrete knowledge use (\cite{Weiss1980}). 

Similarly, journalism uses social science knowledge as \textit{sense-making devices} that can be flexibly applied to interpret news events or contemporary phenomena (\cite{Hallett2019}). Here, we draw on the concept of the \textit{public idea} from the sociology of public social science (\cite{Hallett2019}), which argues that social science ideas become public ideas when research concepts are used by journalism or other mediators that bridge the academy and the public. In this process, social science ideas appear in the news coverage as either the \textit{objects} of the news themselves or the \textit{interpretants} that help make sense of the news. 

While both the journalism and policy domains may draw on social science ideas, we explore whether they tend to utilize the same ideas in different ways. The two-communities theory (\cite{Caplan1979}) noted substantial differences in values, reward systems, and languages between researchers and policymakers, suggesting that they live in two rather separate worlds. Research has also argued that researchers and journalists belong to different social institutions with entirely different professional roles and information functions (\cite{Fjaestad2007}). These institutional differences point to a broader principle: following Luhmann (2013, as cited in \cite{Neuberger2023}), society consists of functionally diﬀerentiated, specialized (sub-)systems, each of which fulﬁlls an exclusive social function and covers a distinctive perspective. These (sub-)systems, or societal domains, produce, validate, disseminate, and appropriate (social science) knowledge in society, with science being considered the highest \textit{epistemic authority} and professional institution that specializes in knowledge generation and supplies other domains with knowledge (\cite{Neuberger2023}). 

Taken together, these perspectives provide the theoretical grounding for our framework: drawing on the enlightenment model and the concept of the public idea, we conceptualize that one of the \textit{indirect} ways knowledge produced by social science research diffuses into other societal domains occurs at the level of social science ideas---that is, named concepts. As these ideas are applied to other societal domains, such as journalism and policy, they are filtered, repackaged, and transformed to tailor them for the respective audiences of these domains. Rather than assuming a direct, instrumental knowledge transfer (typically measured via citations), we view social science knowledge as diffusing circuitously, permeating society over time as a set of conceptual frames, interpretants, and meta-discourses that shape how problems are perceived by actors embedded in different institutions. 

This theoretical framework motivates the empirical relevance for analyzing ideas---conceptualized as named concepts---as trackable units of knowledge diffusion across societal domains. Specifically, we will identify the semantic uptake of social science concepts in each domain, estimate their domain-specific contexts, and track differences across domains and over time. Importantly, we do not assume a directional or causal flow from science into other societal domains. Instead, our framework aims to capture how social science ideas \textit{semantically} circulate and are (re-)contextualized as they move across domains. The resulting conceptual proximity measured via semantic distance across domain-specific sub-corpora is interpreted not as evidence of a single origin or a linear pathway of knowledge transfer but rather as an indicator of the shared conceptual basis that different societal domains interact through \textit{knowledge} (cf. ``knowledge order" (\cite{Neuberger2023})). The remainder of this paper presents a case study of media effects theories as a central part of research ideas from the field of communication science, tracing the diffusion of these prominent social science ideas. 

\section{Methodology}
\subsection{Data}

We collected data from three domains---research, journalism, and policy-making---to study the idea-level diffusion of social science knowledge. Our study focuses specifically on communication science ideas as a particularly relevant case for applying our framework to study idea-level knowledge diffusion. The empirical basis is the media effects theories, which are a central part of communication science ideas. This includes 33 named concepts used to represent 29 theories curated by \cite{Neuman2011}, based on 36 seminal books and articles on media effects research, such as agenda-setting (\cite{McCombs1972}), the two-step flow model (\cite{Katz1957}), and the spiral of silence (\cite{NoelleNeumann1974}). Notably, as noted in the paper (\cite{Neuman2011}), many theoretical traditions that became central to communication science were not originally \textit{effects theories} in the narrow (psychological) sense, but draw on a broad range of sociological, political, or cultural theories in social sciences, e.g., \textit{social network} (\cite{Granovetter1973}) and \textit{public sphere} (\cite{Habermas1991}). Table \ref{tab:media_effects} shows the full list of search terms that we used to retrieve relevant documents.

\begin{table}[h!] 
\centering
\caption{\textbf{Search terms for the media effects theory concepts.} The table shows the list of search strings for retrieving documents mentioning the media effects theory concepts (\cite{Neuman2011}).}
\label{tab:media_effects}

\begin{tabular}{p{12cm}}\toprule
 Search terms\\\midrule

 agenda setting, attitude change, attribution theory, channel effect, cognitive dissonance, computer mediated communication, cultivation theory, differential media exposure, diffusion theory, disposition theory, elaboration likelihood model, framing theory, knowledge gap theory, lasswell linear model, media dependency, media hegemony, minimal effect, parasocial theory, persuasion, priming, public sphere, selective exposure, shannon linear model, social capital, social construction of reality, social identity, social learning, social networks, spiral of silence, third person theory, two step flow, uses and gratifications, voting research \\
\bottomrule
\end{tabular}

\end{table}

\begin{figure}[h]
\centering
\includegraphics[width=\textwidth]{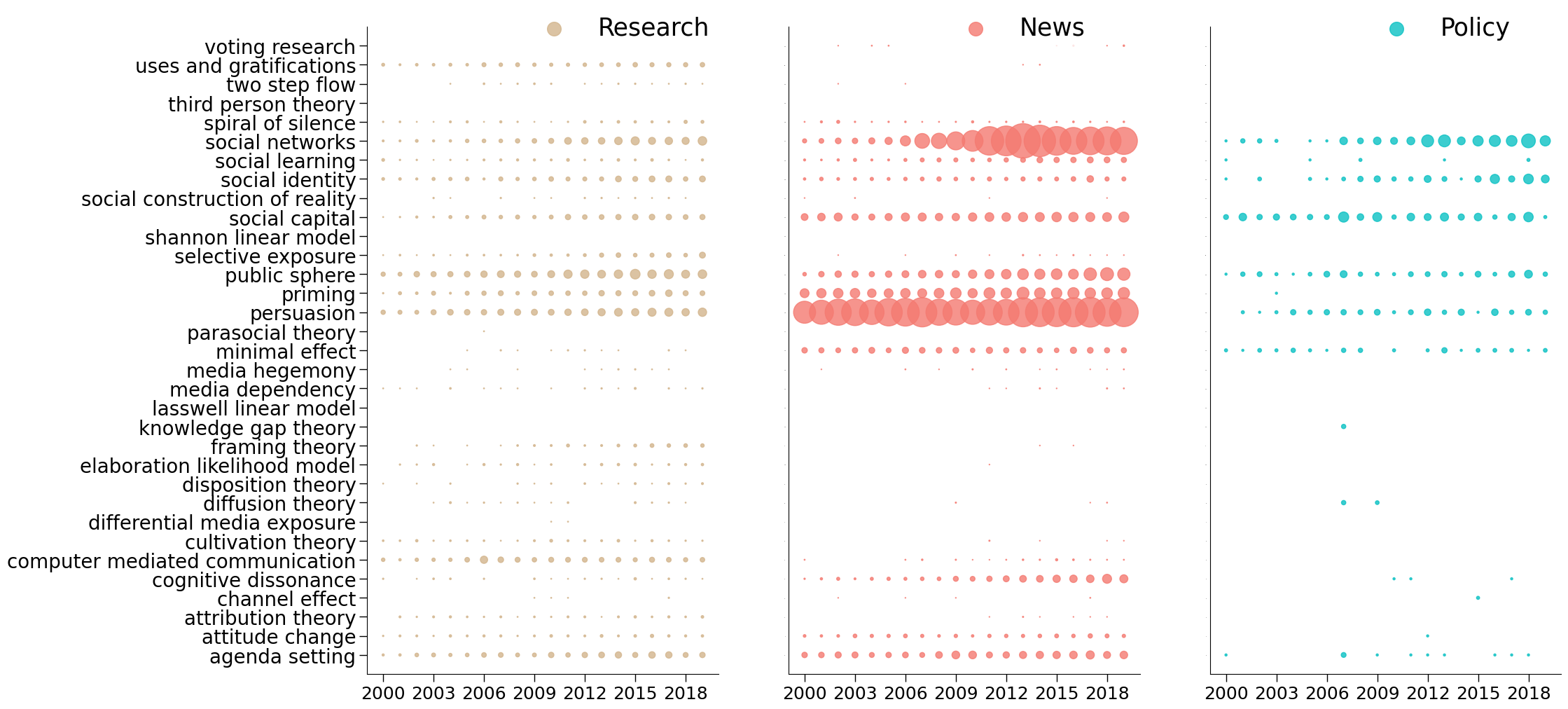}
\caption{\textbf{Prevalence of the media effects theories across three domains.} The graph shows the number of documents mentioning the relevant concepts from 2000 to 2019. Each circle in the graph denotes at least one document in a given year and a specific domain that mentions the concepts. The size of the circle is proportional to the number of documents.}
\label{fig:overall_count}
\end{figure}

We collected a total of 72,703 documents mentioning these concepts, published between 2000 and 2019. We chose 2019 as the endpoint of data collection to avoid COVID-19-related shocks that could lead to substantive shifts and confound our cross-domain comparison. This dataset allows us to track the prevalence of these ideas over time (Figure \ref{fig:overall_count}). Specifically, we used the OpenAlex API to retrieve relevant research papers published in journals indexed in the Web of Science Communication Category. The OpenAlex database provides freely available metadata of scientific entities (e.g., publications, authors, and journals) and their connections (\cite{Priem2022}). The WoS Communication category (2023) contains 245 English journals, 239 (98\%) of which are found in the OpenAlex database. Using a keyword-based approach, we identified 5,505 papers published in 220 journals, each mentioning at least one of the relevant concepts in their titles or abstracts. 

For the journalism domain, we used the Factiva business information database to collect the full texts of relevant English news articles from EU countries and the UK. Specifically, we collected English-language news articles from European newspapers and some of the top UK newspapers\footnote{The selected top UK newspapers include Daily Mail, The Daily Telegraph, Financial Times, The Guardian, The Independent, The Mail on Sunday, The Sunday Telegraph, The Sunday Times, and The Times.}. This resulted in 66,456 news articles that mentioned the relevant concepts. Finally, we collected policy documents from the European Union law and public documents database, EUR-Lex, obtaining 742 EU policy documents mentioning the relevant concepts.

\subsection{Methods}

We use the embedding regression approach (\cite{Rodriguez2023}) to analyze how the use of these ideas varies across different domains and over time. Embedding models, such as GloVe and Transformer (\cite{Pennington2014, Vaswani2017}), are widely used to encode the syntactic and semantic structures in texts. These models intrinsically assume the structuralist view on language that the meaning of words arises from contexts in the text (i.e., word co-occurrences) (\cite{Rodriguez2023}). It aligns well with the distributional hypothesis by Firth (1957)---``you shall know a word by the company it keeps.'' Recently, these embedding models have evolved rapidly and have revolutionized the way textual data is transformed into meaningful measures, not least for mining texts for social theory (\cite{Evans2016}).

The embedding regression approach, specifically, employs an embedding-based strategy, \textit{\`a la carte embeddings}
(ALC) (\cite{Khodak2018}), that computes the representation of the focal word from the additive information of the words in the context window around it (\cite{Rodriguez2023}). For instance, suppose we have the focal word, \textit{apple}, in two sentences; this approach will generate two different embeddings when the two texts mention the word \textit{apple} in two contexts (e.g., food and technology)---here, context is given as several words on each side of the instance of apple. In a nutshell, this approach generates context-specific embeddings for the focal words based on the different contextual words for the focal words across different sub-corpora. 

Technically, we started by applying a range of standard pre-processing techniques to our dataset, including lowercasing the texts and removing punctuation, symbols, numbers, and English stop words. Next, we excluded words with two or fewer characters (e.g., \textit{to}, \textit{it}) and words that appear less than five times. Note that we preserved the position of the removed infrequent words using padding such that non-adjacent words would not become adjacent after processing. Our pre-processing results in 122,735 different tokens and a cumulative count of 40,397,004 tokens. 

After building the corpus, we used standard pre-trained embeddings, which are word vectors that have been fit to some large corpus, and we locally fit them to our corpus, as suggested by (\cite{Rodriguez2023}). Here, we used the glove.6B.300d.txt\footnote{https://www.kaggle.com/datasets/thanakomsn/glove6b300dtxt}, which included 6B tokens and 300-dimensional vectors. Based on the pre-trained word embeddings, we obtain the context-specific embeddings of the media effects concepts using a linear transformation of the average embeddings for the words within the context of six tokens. We focus specifically on the terms that appeared in all three sub-corpora (N=13), including \textit{public sphere},  \textit{persuasion},  \textit{social networks}, \textit{agenda setting}, \textit{social capital}, \textit{channel 
effect},
\textit{attitude change}, \textit{social learning}, \textit{cognitive dissonance}, \textit{social identity}, \textit{minimal 
effect}, \textit{diffusion theory},  and \textit{priming}. Note that we used both the plural and singular forms 
for these terms. We have also trained a domain-specific GloVe (\cite{Pennington2014}) model based on our corpus, with a window size of 6 and an embedding dimension of 300. The results remain largely consistent with the domain-specific embeddings. See the Supplementary Tables \ref{tab:domain}-35 for more details. 

Based on the trained embeddings of focal words (i.e., media effect concepts), we set up a multivariate regression framework, where each observation is an embedding of the media effect concept in the corpus from the studied domains (i.e., research, journalism, and policy). The regression-like framework allows us to make claims about the statistical significance of the differences in embeddings and explore important covariates (i.e., domain). We estimate the following regression: 

\[
Y = \beta_0 + \beta_1\,\text{Corpus} + \beta_2\,\text{Year} + \varepsilon
\]

Specifically, the coefficient $\beta_0$ estimates the embedding of the concept in the research domain as the reference domain, and $\beta_1$ estimates the corpus coefficient matrix, that is, the additional shifts (relative to research) for journalism and policy domains, and we control for the effects of publication year of the documents. Next, we take the Euclidean norms of the coefficients to summarize the domain differences, which measure how different one domain is from the research domain in a relative sense. Importantly, these magnitudes are relative and not directly interpretable, but we can still assess if the semantic difference is statistically significant. Further, we can compare coefficients between journalism and policy to tell which domain deviates more from research. 

\section{Results}
\subsection{The prevalence of communication science ideas and the uptake in other domains}

\begin{figure}
    \centering
    \includegraphics[width=\textwidth]{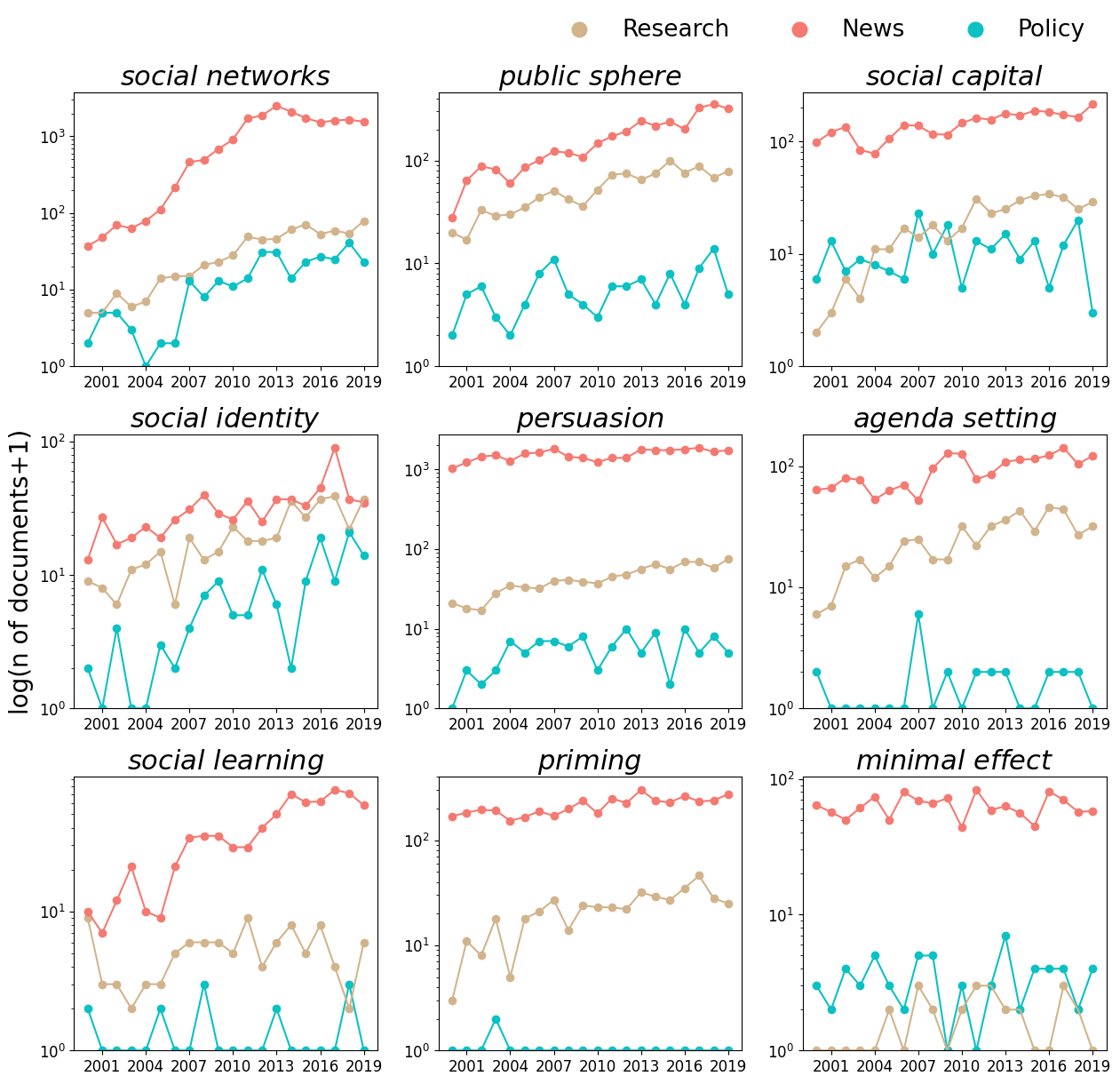}
    \caption{\textbf{Corpus size per concept across domains and time.} Here, the x-axis shows the publication year, and the y-axis shows the number of retrieved documents.}
    \label{fig:9_counts}
\end{figure}

Our analysis first identified mentions of prominent communication science ideas across the three studied domains. As shown in Figure~\ref{fig:overall_count}, multiple named concepts (13 of 33) appear in all three societal domains. Most concepts (29 of 33) are present in the news domain, while only a few concepts (14) appear in policy documents. Table \ref{tab:table2} and Figure \ref{fig:9_counts} provide further details on the number of documents mentioning the concepts that are present in all domains. Closer inspection shows that more popularized and generic concepts, such as \textit{social networks}, \textit{public sphere}, \textit{persuasion}, and \textit{minimal effects}, are frequently mentioned across all domains. Specifically, \textit{public sphere} (with 1,068 documents mentioning this concept), \textit{persuasion} (862), and \textit{social networks} (645) are the most prevalent concepts in the research domain. In the news domain, \textit{persuasion} (30,688), \textit{social networks} (19,430), and \textit{priming} (4,262) are most frequently mentioned. In the context of policy documents, the most mentioned concepts are \textit{social networks} (274), \textit{social identity} (193), and \textit{social capital} (115). Further, we find that several prominent concepts in the research corpus, including \textit{computer-mediated communication} (446 mentions), \textit{uses and gratifications} (264), and \textit{selective exposure} (196), are absent from policy documents. These ideas also receive limited attention in the news domain, with only 48 news articles mentioning \textit{computer-mediated communication}, 18 mentioning \textit{selective exposure}, and 3 mentioning \textit{uses and gratifications}. Supplementary Table \ref{tab:fulltable} presents the number of documents mentioning the 33 named concepts in each domain. 

Our results, while largely descriptive, indicate that the idea-level diffusion of communication science research is highly selective, while some ideas travel to other domain(s) and others do not. Note that we do not intend to provide a causal explanation for the conditions of a successful idea-level diffusion. As noted in previous research (\cite{Hallett2019}), it is nearly impossible to explain or identify a formula for the success of particular social science ideas. This is due to multiple interdependent causes, the role of luck, and a lack of information on negative instances. Therefore, we restrict our analysis to describing the conceptual proximity across domains \textit{after} these ideas appear in other domains. In the following section, we focus specifically on the concepts that occur in all three sub-corpora and examine their domain-specific uptakes and how these evolve over time.

\begin{table}[h!] 
    \centering
        \caption{\textbf{The number of documents mentioning the studied concepts}}
    \begin{tabular}{cccc}
    \hline
     Concepts    &  Research &  News & Policy \\
     \hline
     public sphere     & 1068  &  3252& 96\\
    persuasion     &  862&  30688& 92\\
    social networks     & 645 &  19430& 274\\
    agenda setting      &  478&  1848&13 \\
    priming     & 419 &  4262& 1\\
    social identity      & 370  & 625  & 115\\
    social capital      &  358&  2846& 193\\
    attitude change      &  101&  469& 1\\
    social learning      &  83&  719& 7\\
    cognitive dissonance      & 34  & 1245  &3 \\
    diffusion theory        &  25&  6& 7\\
     minimal effect         &  14&  1239& 47\\
    channel effect                & 5 & 5 & 2 \\
    \hline
    \end{tabular}
    \label{tab:table2}
\end{table}

\subsection{Embedding regression}
We used an embedding regression approach \parencite{Rodriguez2023} to measure the context-specific use of these concepts. While most existing word embedding-based approaches are used for descriptive or predictive purposes, this method enables us to draw statistical inferences. In particular, it allows us to compare different instances of a concept in an embedding space as a function of domain (i.e., research, journalism, and policy-making) while controlling for other covariates such as the publication year of the document. 

Tables~\ref{tab:regression} and~\ref{tab:regression2} present the results of the embedding regression for the studied concepts. 
Due to the limited number of instances in the policy documents, the models encountered computationally singular issues for some concepts and failed to produce reliable estimates. Hence, we omitted those concepts and focused on the concepts when the models converged. The reported estimates are the Euclidean norms of the domain coefficients, which provide a single scalar measure of how large or small the semantic shift is from the research domain (i.e., normed domain effects). We find that all reported coefficients are statistically significant, indicating that the use of these concepts differs significantly across domains. In other words, the regression analysis shows a measurable semantic shift when these concepts are used outside academia. Further, what stands out in the regression tables is that the normed domain effects of policy are larger than news for most concepts (e.g., \textit{public sphere}, \textit{agenda setting}, and \textit{social identity}), with only two exceptions of \textit{social networks} and \textit{social capital}. That is, on average, policy use usually deviates farther from research use than news. This suggests that research and news share more similar understandings of the concepts than policy in most cases.


\begin{table}[H]
\centering
\caption{\textbf{Regression estimates}}
\begin{tabular}{l c c c c }
\hline
 & Social networks & Public sphere & Social capital & Agenda Setting  \\
\hline
News             & 4.632 (0.096)  & 4.940 (0.084) & 5.575 (0.109) &6.955 (0.110)  \\
Policy           & 4.478 (0.153)  & 6.203 (0.237) & 5.500 (0.170) &9.698 (0.665)  \\
Publication year & 0.240 (0.005) & 0.150 (0.006) & 0.190 (0.007) & 0.223 (0.013)  \\N & 26678 & 4882 & 4716& 2819\\
\hline
\end{tabular}
\\[1ex]
\small Notes: Normed coefficients from regression models; standard errors in parentheses, $p<0.001$. N denotes instances of the focal term across all domains, not the number of documents.  
\label{tab:regression}
\end{table}

\begin{table}[H]
\centering
\caption{\textbf{Regression estimates}}
\begin{tabular}{l c c c c}
\hline
 & Social identity & Persuasion & Social learning & Minimal effect  \\
\hline
News             & 6.873 (0.238)  & 7.106 (0.093) & 7.819 (0.384) & 11.615 (1.167) \\
Policy           & 21.603 (0.561) & 9.170 (0.587) & 14.799 (1.269) & 14.261 (1.162) \\
Publication year & 0.338 (0.022)  & 0.070 (0.002) & 0.327 (0.019) & 0.238 (0.015) \\
N & 1268 & 31113 & 929 & 1223 \\
\hline
\end{tabular}
\\[1ex]
\small Notes: Normed coefficients from regression models; standard errors in parentheses, $p<0.001$. N denotes instances of the focal term across all domains, not the number of documents.  
\label{tab:regression2}
\end{table}

The embedding regression approach also produces semantically interpretable \textit{nearest neighbors} of the focal words. It does so by calculating the cosine similarity between the ALC embedding of the concepts and the pre-trained embeddings of other tokens. For each concept, we obtain three groups of the nearest neighbors in the three domains, respectively. This allows us to examine \textit{qualitatively} how the understanding of the concept shifts across domains. We provide the full lists of the top 10 nearest neighbors for the concepts in each domain in Supplementary Tables \ref{tab:sc}-\ref{tab:diffusion}. Tables \ref{tab:sn} and \ref{tab:publicsphere} show the results for two examples---\textit{social networks} and  \textit{public sphere}. 

After examining these semantic neighbors of the concepts in each domain, we find that the use of these concepts largely shifts contexts across domains. Specifically, terms such as \textit{theories}, \textit{context}, \textit{concepts} frequently co-occur with most concepts in the research domain, including \textit{social networks}, \textit{public sphere}, \textit{social identity}, \textit{social learning}, \textit{agenda setting}, and \textit{cognitive dissonance}. In the news articles, the nearest neighbors indicate a wide variety of \textit{social} contexts, including \textit{politics}, \textit{religion}, \textit{culture}, and \textit{teaching}. In the policy documents, the neighbors are more actionable and institutional, such as \textit{participation}, \textit{development}, \textit{intervention}, \textit{skill}, and \textit{competence}, as well as verbs like \textit{enhance}, \textit{promote}, and \textit{overcome}. 

In the case of the two exceptions of concepts (i.e., \textit{social networks} and \textit{social capital}), we find their policy use is comparably closer to the research use. For \textit{social networks}, the research domain is close to terms like \textit{interaction} and \textit{relationships}, the news domain shifts towards \textit{users} and social media (e.g., Facebook, Twitter, and YouTube), while the policy domain emphasizes \textit{websites}, \textit{internet}, and \textit{networking}. Hence, the policy use is relatively more consistent with the research framing compared to the news, as it relates more to the broader structural and communicative aspects (not only social media platforms). Similarly, for social capital, we find that news use emphasizes terms like \textit{community} and \textit{benefit}, whereas policy contexts focus on \textit{development} and \textit{interaction}, which are slightly more strongly linked to the research use (e.g., \textit{interaction} and \textit{relationships}). 
Additionally, our analysis reveals some polysemantic terms, whose uptake in other domains is rather semantically distant from their research use. \textit{Priming}, for instance, was mainly used in the sense of liquid \textit{pump} technology in the news domain. Similarly, the uptake of \textit{persuasion} in the news domain is close to as colloquial terms, such as \textit{whatever} and \textit{kind}.

Taken together, this suggests that the ideas largely shift roles when traveling outside academia---from being the theories themselves in research to sense-making in news to more applied, administrative use in policy. Having discussed the overall picture of the idea-level diffusion of media effect theories based on the conceptual proximity, this paper will next explore how these patterns evolve over time.

\begin{table}[ht]
\centering
\caption{\textbf{The most similar terms to \textit{social networks}}}
\begin{tabular}{l c l c l c}
\hline
\multicolumn{2}{c}{Research} & \multicolumn{2}{c}{News} & \multicolumn{2}{c}{Policy} \\
Term & Similarity & Term & Similarity & Term & Similarity \\
\hline
interaction    & 0.677 & facebook   & 0.769 & websites      & 0.686 \\
interactions   & 0.646 & users      & 0.726 & web           & 0.640 \\
relationships  & 0.621 & internet   & 0.725 & online        & 0.635 \\
communication  & 0.543 & web        & 0.716 & internet      & 0.632 \\
contexts       & 0.540 & twitter    & 0.706 & networking    & 0.631 \\
interpersonal  & 0.538 & online     & 0.685 & blogs         & 0.625 \\
context        & 0.513 & websites   & 0.673 & facebook      & 0.622 \\
phenomena      & 0.505 & blogs      & 0.633 & forums        & 0.607 \\
concepts       & 0.501 & myspace    & 0.621 & user          & 0.594 \\
collaborative  & 0.496 & youtube    & 0.603 & twitter       & 0.589 \\
\hline
\end{tabular}
\label{tab:sn}
\end{table}

\begin{table}[ht]
\centering
\caption{\textbf{The most similar terms to \textit{public sphere}}}
\begin{tabular}{l c l c l c}
\hline
\multicolumn{2}{c}{Research} & \multicolumn{2}{c}{News} & \multicolumn{2}{c}{Policy} \\
Term & Similarity & Term & Similarity & Term & Similarity \\
\hline
discourse      & 0.731 & politics     & 0.663 & participation  & 0.624 \\
perspectives   & 0.567 & religion     & 0.662 & promote        & 0.548 \\
context        & 0.549 & religious    & 0.612 & participate    & 0.538 \\
theories       & 0.521 & regard       & 0.602 & citizens       & 0.534 \\
philosophical  & 0.518 & discourse    & 0.591 & governmental   & 0.518 \\
theory         & 0.518 & debate       & 0.586 & organizations  & 0.515 \\
sociological   & 0.518 & notion       & 0.581 & promoting      & 0.507 \\
contexts       & 0.513 & political    & 0.575 & institutions   & 0.506 \\
concepts       & 0.509 & indeed       & 0.574 & encourage      & 0.504 \\
sphere         & 0.497 & faith        & 0.574 & involvement    & 0.504 \\
\hline
\end{tabular}
\label{tab:publicsphere}
\end{table}

\subsection{Temporal dynamics}
\begin{figure}
    \centering
    \includegraphics[width=\textwidth]{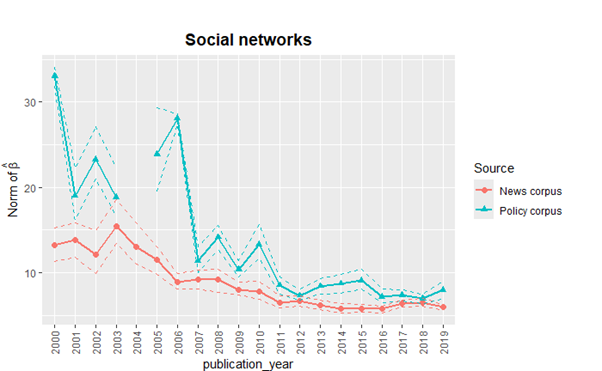}
    \caption{\textbf{Embedding-based distance across domains and time} Here, the x-axis shows the publication year, and the y-axis shows the normed $\beta$.}
    \label{fig:sm_years}
\end{figure}

\begin{figure}
    \centering
    \includegraphics[width=\textwidth]{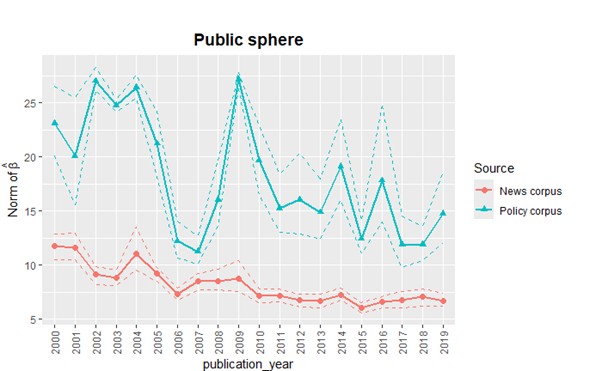}
    \caption{\textbf{Embedding-based distance across domains and time} Here, the x-axis shows the publication year, and the y-axis shows the normed $\beta$.}
    \label{fig:ps_years}
\end{figure}

We further estimated the normed domain effects, the normed $\beta$, for every year from 2000 to 2019. When the normed $\beta$ drops, the use of the concept is becoming more similar between the reference domain (i.e., research domain) and the journalism or policy domain. Due to the limited number of policy instances in certain years, we were unable to estimate domain effects for policy in some periods. The time series of $\beta$s for each concept is shown in the Supplementary Figures \ref{fig:sc_years}-\ref{fig:as_years}. Figures \ref{fig:sm_years} and \ref{fig:ps_years} plot the time series for two examples---\textit{social networks} and  \textit{public sphere}. 
We find that throughout the observation period, the normed domain effects for policy are generally larger than for news across all concepts. This indicates that news use of these concepts is consistently more similar to the research than the policy use. When looking at the changes over time, we observe a notable convergence for two concepts---\textit{social networks} and \textit{social capital}---whose normed domain effects generally decline over time, suggesting that their use is becoming more similar to the research use. Further, the policy effects are getting closer to the news effects over time for these two concepts. For other concepts, we do not observe a clear convergence pattern. \textit{Public sphere}, for example, shows a fluctuating trend, with the policy effects rising and falling over time and consistently larger than the news effects. Note that the magnitudes of the normed coefficients are relative and not directly interpretable as substantive distances, and we focus only on their comparative effect size.

To understand the substance of the change, we examine the semantic neighbors of the concepts in each domain in each year. For details, see Supplementary Tables \ref{tab:socnet_yearly}-\ref{tab:agendasetting_yearly}. 
In the case of \textit{social networks}, we find that nearest neighbors remain mostly theoretical and analytical terms in the research domain, focusing on \textit{interaction}, \textit{relationships}, and \textit{communication} (without any platform names). In the news context, early years focus more on the social and communal terms (e.g., \textit{relationships}, \textit{community}, and \textit{friends}). Since 2006, the news use turned to \textit{online} and \textit{digital} contexts, and later it focused mostly on social media platforms. Since 2018, the news use has broadened to functions and concerns brought by platformization (\textit{protect}, \textit{allow}, \textit{need}). In the policy context, it was initially closely related to \textit{facilitate} social cohesion (e.g., \textit{cohesion}, \textit{facilitate}, \textit{intergenerational}, and \textit{territorial}). Around 2007, the policy use turned to \textit{knowledge}, \textit{tools}, \textit{skills}, and later \textit{literacy}. From 2008 onward, the neighbors became more user- and ICT-oriented (e.g., \textit{user}, \textit{ict}, and \textit{internet}). Specific \textit{platform} mentions appeared later. By the end of the period, terms such as \textit{websites}, \textit{platform}, and \textit{organisations} are prevalent, reflecting tangible infrastructures and programs in digital policy. 

A similar pattern in policy use was observed for \textit{social capital}. In the policy document, early neighbors focused on equity (e.g., \textit{women}, \textit{equal}), \textit{local} \textit{organization}, and \textit{human} \textit{development}). This was accompanied by impact and programmatic language (e.g., \textit{improve}, \textit{impact}, \textit{importance}). From 2014, it has turned to \textit{growth}, \textit{economic}, \textit{improvement}, and \textit{inequality}. From 2018 onward, it shifts towards \textit{entrepreneurship}, \textit{governance}, and \textit{capital}. This suggests that the policy use is increasingly administrative and programmatic, which can be mapped into development programs and institutional initiatives with measurable outcomes. By contrast, the news use mainly shifted from community and relationship topics (e.g., \textit{community}, \textit{trust}, and \textit{social}) to economic and financial topics (e.g., \textit{wealth}, \textit{benefit}, and \textit{investment}) over time. In the research documents, the focus generally remains on \textit{relationships} and \textit{interactions} throughout the studied period. Hence, we consider these two concepts to be relatively more \textit{practically oriented} as their policy neighbors reference tangible infrastructures or administrative actions---potentially indicating policy instruments and programs. 

In striking contrast, other concepts do not show such a level of specificity regarding policy instruments. \textit{Public sphere}, for instance, tends to co-occur with broad terms with normative implications, such as \textit{citizens}, \textit{democracy}, and \textit{participation}. It shifted from \textit{gender}, \textit{racism}, and \textit{international} topics to \textit{youth}, \textit{institution}, and \textit{government}, focusing on \textit{equality} and \textit{participation}. Even if some terms denote the aims and procedures (e.g., \textit{promote}, \textit{encourage}, and \textit{ongoing}), this reflects more of a discursive orientation around equality and participation, which is less tangible regarding the policy instrument. A similar pattern is observed for \textit{social identity}. In policy documents, its nearest neighbors remain biological and psychological terms, including \textit{physical}, \textit{mental}, \textit{physiological}, and \textit{psychological}. This likely reflects a descriptive or diagnostic use rather than policy instruments or programmatic actions. With regard to domains, we find that the research use remains theoretical and analytical for both \textit{public sphere} and \textit{social identity} throughout the studied period. 
By contrast, news use spans a broad range of political, cultural, and social contexts, where the concepts often function as flexible interpretive devices. In practice, the same concept is adapted to fit event-specific frames. For example, news mentions of \textit{public sphere} co-occur with terms such as \textit{politics}, \textit{ideology}, \textit{democracy}, as well as \textit{religion}, \textit{culture}, \textit{faith}, and \textit{morality}. Similarly, news use of \textit{social identity} appears across heterogeneous settings—co-occurring with \textit{unique}, \textit{sexual}, \textit{genetic}, and \textit{british}, \textit{islamic} topics, alongside interpersonal and cultural frames like \textit{friendship}, \textit{portraits}, \textit{belong}ing, \textit{value}, \textit{cultures}, and \textit{history}.

Taken together, \textit{social networks} and \textit{social capital} are relatively more practically oriented than other concepts in this study. We find that their nearest neighbors in the policy domain link to more practically-oriented terms and reflect more tangible infrastructures and programs. This corresponds to the early results from the regression analysis as shown in Tables \ref{tab:sn}---indicating their policy use is closer to research use compared to news use. By contrast, other concepts are more used as interpretive sense-making devices (e.g., \textit{public sphere} and \textit{social identity}). They were used across a variety of contexts, especially in the news as interpretants, suggesting that they are flexible and compatible enough to various (social) settings, but less actionable in policy terms.

\section{Discussion and conclusion}
This paper has developed a novel measurement framework to study cross-domain knowledge diffusion beyond citations. By focusing on named social science concepts as the unit of analysis, it complements and expands the existing citation- and mention-based measures with a view of knowledge diffusion that attends to meaning and context. Grounded in literature from political science, sociology, and communication science, we connect the scientometric measurements to thoughts and theories that emphasize the \textit{conceptual} use of social science knowledge in other domains. Specifically, we conceptualize that one of the indirect pathways through which social science knowledge diffuses into other societal domains occurs at the level of social science ideas---that is, named concepts. These ideas permeate society over a longer period of time as a set of conceptual frames, interpretants, and meta-discourses that potentially shape how actors perceive social problems. 

To operationalize this framework, we use an embedding regression approach, which allows us to compare contextualized meanings across domains and draw statistical inferences. We estimate normed domain effects (i.e., how journalism or policy use deviates from research use for each concept) and use nearest neighbors to interpret the semantic meaning of use (i.e., (re-)contextualization of the concepts across domains). Empirically, we focus on media effect theory concepts identified by Neuman and Guggenheim (\citeyear{Neuman2011}), which incorporate theoretical traditions that have become central to communication science. This is a particularly relevant case for studying idea-level knowledge diffusion, as it encompasses a broad range of psychological, sociological, political, or cultural theories in social sciences.  We collected a large dataset of 72,703 documents (2000-2019) from the research, journalism, and policy domains that mention 33 named media-effects theory concepts. 

Our results indicate a largely heterogeneous and complex idea-level diffusion, during which some ideas are more likely to travel from the academic to other domains than others. When they do, we observe that usage in the policy domain usually deviates further from research than in the news domain, and ideas often shift roles across domains—from being the theories themselves in research to sense-making in news to applied, administrative use in policy. Informed by the nuances in our analysis, we characterize three broad groups of ideas: practically oriented (e.g., social networks, social capital), interpretive (e.g., public sphere, social identity), and polysemes (e.g., priming, persuasion). These groups differ meaningfully in their patterns of cross-domain conceptual proximity and qualitatively in how their semantic neighbors evolve over time.

Overall, this study offers a potentially generalizable measurement framework for future research on idea-level knowledge diffusion and could motivate further (meta-)theoretical explorations of how various societal domains interact through knowledge. Moreover, our findings demonstrate the societal impact of communication science in journalism and policy-making. Further, we reveal the different uptake of social-science research in these two domains. We believe these contributions advance our understanding of inherently qualitative predictors of knowledge transfer and research impact, adding to traditional publication metrics such as citation count (\cite{Garfield1955}) and \textit{h}-index (\cite{Hirsch2005}).

Our study has several important limitations. First, the use of the term \textit{knowledge diffusion} may imply causal or temporal dynamics. However, we have made it clear that our approach does not assume a directional or causal flow from research to other societal domains. We operationalize knowledge diffusion \textit{descriptively} to trace how named social science concepts appear and evolve across domains. Future studies could extend our work to identify directional influence or mechanisms, or model the temporal sequences, such as feedback loops, that establish and reinforce the relevance of social science research in other societal domains.

Second, the embedding regression approach has important methodological constraints. As noted by the authors (\cite{Rodriguez2023}), estimates can be biased for rare terms and when comparing groups with substantially different sample sizes (as is the case for policy mentions here). Furthermore, we note that the Euclidean norm of regression coefficients is inherently non-negative and should be read as a measure of magnitude rather than a directly interpretable distance. 

Third, our policy corpus is smaller than the other domains and is limited to EU legislative texts. Our sample is also restricted to English-language texts, which limits its generalizability to non-English contexts, especially for the news and policy domains, where the local contexts might yield interesting results. Finally, our case study focuses on (interdisciplinary) communication science, which spans a broad range of disciplines in the social sciences and humanities. Future work could extend this study to other social science fields and disciplines, producing a more comprehensive picture of how social science knowledge diffuses \textit{conceptually}.

Notwithstanding these limitations, this study aligns measurement with how social science knowledge diffuses to other societal domains in a broader sense: less as a direct reference to specific works or an instrumental use for policy change, more as a set of conceptual frames that accumulate in public discourse through a gradual, indirect process. Through recognizing such diffusion, we gain a more comprehensive understanding and a more positive view of the societal impact of social science research (\cite{Weiss1980}). In this sense, our study takes a step toward measuring the societal impact of research beyond citations. 

\section{\textbf{Acknowledgments}}

We thank Merja Mahrt for suggestions on the selection of communication science concepts. We thank Alexandra Malaga, colleagues, and audience members at the \textit{Bridging Approaches in the Sciences Studying Science} Workshop (Munich, 2025) for helpful comments. We further thank our student assistants Joana Becker, Carolin Stock, and Carlo Uhl for their help during the data collection. This project was funded by the seed funding (2023) from the Weizenbaum Institute (WI), Berlin. We would like to acknowledge funding by the Federal Ministry of Education and Research of Germany (BMBF) under grants No. 16DII131 and 16DII135 (Weizenbaum-Institut für die vernetzte Gesellschaft – Das Deutsche Internet-Institut). 

\section{\textbf{Data availability}}

The dataset and the code have been published in an online \href{https://github.com/YangliuF95/knowledge_diffusion}{repository}. Note that we are unable to provide the full-text news articles due to Factiva’s proprietary licensing restrictions. 

\section{\textbf{Author contributions}}
All authors conceived the project and designed the study; Y.F. and K.B. collected and analyzed the data; Y.F. wrote the manuscript; all authors edited the manuscript.

\printbibliography
\clearpage
\appendix
\section*{Supplementary Figures}

\begin{figure}[H]
    \centering
    \includegraphics[width=\textwidth]{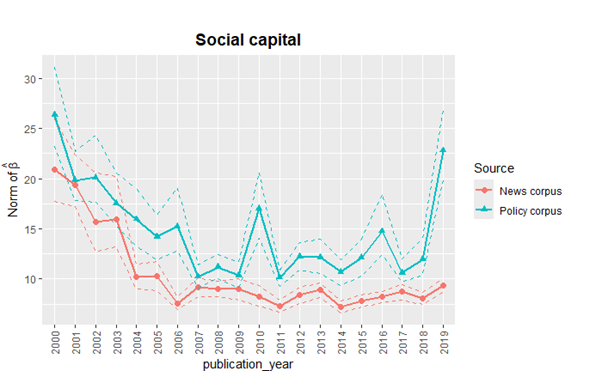}
    \caption{\textbf{Embedding-based distance across domains and time} Here, the x-axis shows the publication year, and the y-axis shows the normed  $\beta$.}
    \label{fig:sc_years}
\end{figure}

\begin{figure}[H]
    \centering
    \includegraphics[width=\textwidth]{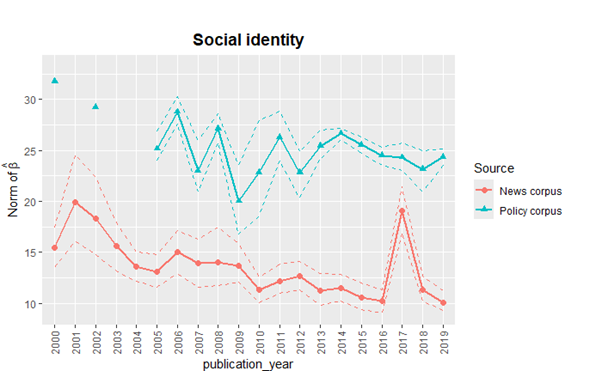}
    \caption{\textbf{Embedding-based distance across domains and time} Here, the x-axis shows the publication year, and the y-axis shows the normed  $\beta$.}
    \label{fig:si_years}
\end{figure}
\begin{figure}[H]
    \centering
    \includegraphics[width=\textwidth]{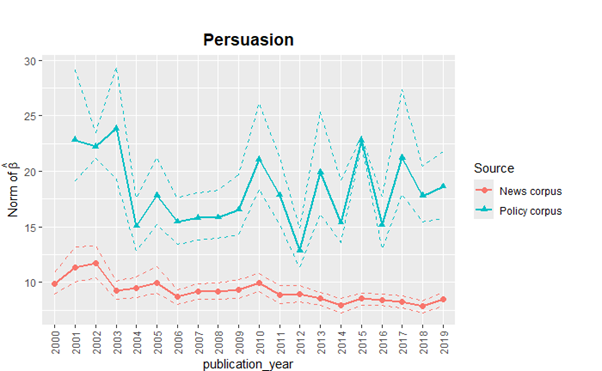}
    \caption{\textbf{Embedding-based distance across domains and time} Here, the x-axis shows the publication year, and the y-axis shows the normed  $\beta$.}
    \label{fig:persuasion_years}
\end{figure}
\begin{figure}[H]
    \centering
    \includegraphics[width=\textwidth]{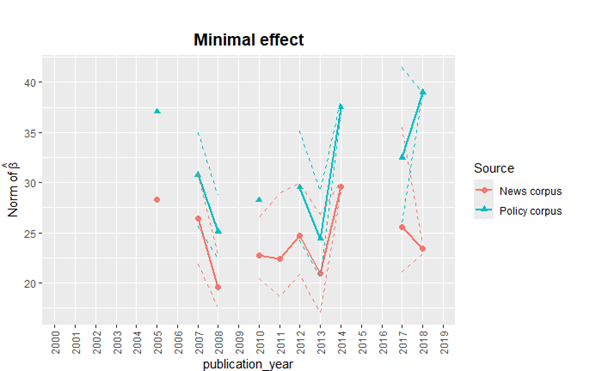}
    \caption{\textbf{Embedding-based distance across domains and time} Here, the x-axis shows the publication year, and the y-axis shows the normed  $\beta$.}
    \label{fig:me_years}
\end{figure}
\begin{figure}[H]
    \centering
    \includegraphics[width=\textwidth]{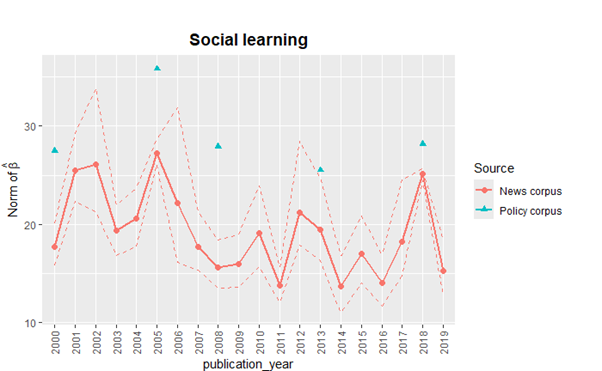}
    \caption{\textbf{Embedding-based distance across domains and time} Here, the x-axis shows the publication year, and the y-axis shows the normed  $\beta$.}
    \label{fig:sl_years}
\end{figure}
\begin{figure}[H]
    \centering
    \includegraphics[width=\textwidth]{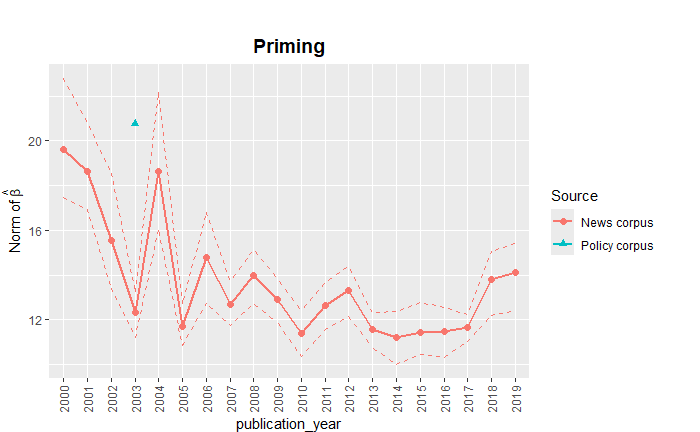}
    \caption{\textbf{Embedding-based distance across domains and time} Here, the x-axis shows the publication year, and the y-axis shows the normed  $\beta$.}
    \label{fig:priming_years}
\end{figure}
\begin{figure}[H]
    \centering
    \includegraphics[width=\textwidth]{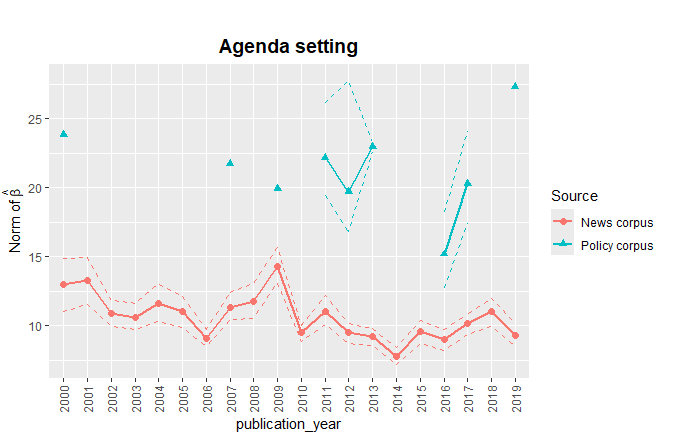}
    \caption{\textbf{Embedding-based distance across domains and time} Here, the x-axis shows the publication year, and the y-axis shows the normed  $\beta$.}
    \label{fig:as_years}
\end{figure}

\section*{Supplementary Tables}

\begin{table}[htbp]
\centering
\caption{The number of documents mentioning media effect theories across domains}
\begin{tabular}{lrrr}
\hline
\textbf{Concept} & \textbf{Research} & \textbf{News} & \textbf{Policy} \\
\hline
public sphere & 1068 & 3252 & 96 \\
persuasion & 862 & 30688 & 92 \\
social networks & 645 & 19430 & 274 \\
agenda setting & 478 & 1848 & 13 \\
computer mediated communication & 446 & 48 & --- \\
priming & 419 & 4262 & 1 \\
social identity & 370 & 625 & 115 \\
social capital & 358 & 2846 & 193 \\
uses and gratifications & 264 & 3 & --- \\
selective exposure & 196 & 18 & --- \\
framing theory & 104 & 2 & --- \\
attitude change & 101 & 469 & 1 \\
social learning & 83 & 719 & 7 \\
spiral of silence & 81 & 101 & --- \\
cultivation theory & 71 & 6 & --- \\
attribution theory & 67 & 8 & --- \\
elaboration likelihood model & 66 & 1 & --- \\
cognitive dissonance & 34 & 1245 & 3 \\
disposition theory & 29 & --- & --- \\
diffusion theory & 25 & 6 & 7 \\
media dependency & 24 & 11 & --- \\
two step flow & 23 & 2 & --- \\
social construction of reality & 18 & 5 & --- \\
minimal effect & 14 & 1239 & 47 \\
media hegemony & 11 & 16 & --- \\
channel effect & 5 & 5 & 2 \\
differential media exposure & 2 & --- & --- \\
parasocial theory & 1 & --- & --- \\
knowledge gap theory & --- & --- & 4 \\
voting research & --- & 16 & --- \\
\hline
\end{tabular}
\label{tab:fulltable}
\end{table}

\begin{table}[h!]
\centering
\caption{The most similar terms to \textit{social capital}.}
\begin{tabular}{l c l c l c}
\hline
\multicolumn{2}{c}{Research} & \multicolumn{2}{c}{News} & \multicolumn{2}{c}{Policy} \\
Term & Similarity & Term & Similarity & Term & Similarity \\
\hline
interaction     & 0.651 & communities  & 0.633 & development  & 0.525 \\
relationships   & 0.619 & community    & 0.623 & enhance      & 0.516 \\
interactions    & 0.614 & benefit      & 0.577 & social       & 0.495 \\
interpersonal   & 0.565 & prosperity   & 0.573 & interaction  & 0.493 \\
relationship    & 0.537 & build        & 0.571 & fostering    & 0.481 \\
understanding   & 0.521 & wealth       & 0.571 & economic     & 0.474 \\
context         & 0.515 & social       & 0.567 & cohesion     & 0.472 \\
cultural        & 0.510 & trust        & 0.559 & linkages     & 0.470 \\
relation        & 0.500 & cultural     & 0.558 & human        & 0.467 \\
social          & 0.498 & promote      & 0.557 & strengthening& 0.463 \\
\hline
\end{tabular}
\label{tab:sc}
\end{table}

\begin{table}[H]
\centering
\caption{The most similar terms to \textit{social identity}.}
\begin{tabular}{l c l c l c}
\hline
\multicolumn{2}{c}{Research} & \multicolumn{2}{c}{News} & \multicolumn{2}{c}{Policy} \\
Term & Similarity & Term & Similarity & Term & Similarity \\
\hline
theory        & 0.716 & distinct     & 0.567 & physical       & 0.663 \\
theories      & 0.678 & identity     & 0.558 & mental         & 0.592 \\
hypothesis    & 0.603 & cultural     & 0.530 & physiological  & 0.560 \\
concepts      & 0.584 & religion     & 0.523 & psychological  & 0.529 \\
linguistic    & 0.551 & characteristics & 0.519 & emotional   & 0.487 \\
methodology   & 0.540 & particular   & 0.506 & cognitive      & 0.457 \\
empirical     & 0.533 & culture      & 0.503 & cultural       & 0.443 \\
posits        & 0.523 & cultures     & 0.494 & behavioral     & 0.427 \\
theoretical   & 0.518 & distinguishing & 0.491 & linguistic   & 0.411 \\
interactions  & 0.516 & unique       & 0.490 & behavioural    & 0.404 \\
\hline
\end{tabular}
\end{table}

\begin{table}[H]
\centering
\caption{The most similar terms to \textit{persuasion}.}
\begin{tabular}{l c l c l c}
\hline
\multicolumn{2}{c}{Research} & \multicolumn{2}{c}{News} & \multicolumn{2}{c}{Policy} \\
Term & Similarity & Term & Similarity & Term & Similarity \\
\hline
theory        & 0.612 & whatever   & 0.669 & necessary   & 0.528 \\
discourse     & 0.604 & kind       & 0.662 & sufficient  & 0.517 \\
theories      & 0.602 & give       & 0.650 & determine   & 0.502 \\
concepts      & 0.571 & sort       & 0.642 & obtain      & 0.501 \\
narrative     & 0.561 & want       & 0.639 & seek        & 0.500 \\
reasoning     & 0.556 & enough     & 0.638 & determined  & 0.500 \\
cognitive     & 0.551 & rather     & 0.638 & required    & 0.495 \\
empirical     & 0.543 & take       & 0.637 & appropriate & 0.483 \\
interpersonal & 0.540 & always     & 0.635 & degree      & 0.476 \\
context       & 0.532 & even       & 0.630 & punishment  & 0.471 \\
\hline
\end{tabular}
\end{table}

\begin{table}[H]
\centering
\caption{The most similar terms to \textit{minimal effect}.}
\begin{tabular}{l c l c l c}
\hline
\multicolumn{2}{c}{Research} & \multicolumn{2}{c}{News} & \multicolumn{2}{c}{Policy} \\
Term & Similarity & Term & Similarity & Term & Similarity \\
\hline
suggest      & 0.529 & increase   & 0.644 & gases       & 0.475 \\
era          & 0.524 & expected   & 0.616 & drying      & 0.459 \\
effects      & 0.474 & although   & 0.593 & chemical    & 0.411 \\
suggesting   & 0.461 & affected   & 0.590 & greenhouse  & 0.391 \\
early        & 0.455 & adding     & 0.585 & dioxide     & 0.383 \\
changes      & 0.447 & cost       & 0.584 & warming     & 0.382 \\
emerged      & 0.443 & though     & 0.583 & effect      & 0.380 \\
similar      & 0.443 & affect     & 0.579 & temperature & 0.377 \\
suggests     & 0.438 & result     & 0.578 & cooling     & 0.373 \\
indicate     & 0.436 & said       & 0.573 & amounts     & 0.372 \\
\hline
\end{tabular}
\end{table}

\begin{table}[H]
\centering
\caption{The most similar terms to \textit{social learning}.}
\begin{tabular}{l c l c l c}
\hline
\multicolumn{2}{c}{Research} & \multicolumn{2}{c}{News} & \multicolumn{2}{c}{Policy} \\
Term & Similarity & Term & Similarity & Term & Similarity \\
\hline
theory        & 0.721 & learning    & 0.678 & competence   & 0.635 \\
theories      & 0.689 & teaching    & 0.616 & competences  & 0.535 \\
concepts      & 0.620 & classroom   & 0.608 & interpersonal& 0.527 \\
evolutionary  & 0.600 & classrooms  & 0.581 & creativity   & 0.491 \\
cognitive     & 0.580 & unique      & 0.542 & abilities    & 0.474 \\
phenomena     & 0.572 & spaces      & 0.540 & skills       & 0.465 \\
evolution     & 0.570 & students    & 0.536 & empathy      & 0.450 \\
interaction   & 0.551 & study       & 0.530 & skill        & 0.446 \\
theoretical   & 0.550 & science     & 0.527 & personal     & 0.414 \\
interactions  & 0.549 & space       & 0.514 & experiential & 0.403 \\
\hline
\end{tabular}
\end{table}

\begin{table}[H]
\centering
\caption{The most similar terms to \textit{priming}.}
\begin{tabular}{l c l c l c}
\hline
\multicolumn{2}{c}{Research} & \multicolumn{2}{c}{News} & \multicolumn{2}{c}{Policy} \\
Term & Similarity & Term & Similarity & Term & Similarity \\
\hline
morphological & 0.596 & pump      & 0.789 & promoted       & 0.393 \\
lexical       & 0.595 & pumping   & 0.577 & partnership    & 0.368 \\
semantic      & 0.586 & pumps     & 0.546 & engineering    & 0.319 \\
similarity    & 0.581 & pumped    & 0.521 & promotion      & 0.305 \\
syntactic     & 0.576 & needed    & 0.502 & fund           & 0.297 \\
grammatical   & 0.569 & stimulate & 0.498 & touted         & 0.278 \\
perceptual    & 0.539 & stimulus  & 0.488 & jointly        & 0.276 \\
phonological  & 0.537 & money     & 0.486 & entrepreneurship& 0.271 \\
meanings      & 0.529 & help      & 0.477 & scientifically & 0.267 \\
verb          & 0.523 & boost     & 0.469 & committee      & 0.263 \\
\hline
\end{tabular}
\end{table}

\begin{table}[H]
\centering
\caption{The most similar terms to \textit{agenda setting}.}
\begin{tabular}{l c l c l c}
\hline
\multicolumn{2}{c}{Research} & \multicolumn{2}{c}{News} & \multicolumn{2}{c}{Policy} \\
Term & Similarity & Term & Similarity & Term & Similarity \\
\hline
theories     & 0.579 & magazine    & 0.612 & collaborative  & 0.403 \\
theory       & 0.572 & news        & 0.599 & joint          & 0.402 \\
analysis     & 0.555 & weekly      & 0.579 & participation  & 0.392 \\
context      & 0.547 & commentary  & 0.564 & collaboration  & 0.370 \\
sociological & 0.516 & newspaper   & 0.555 & partnership    & 0.366 \\
focuses      & 0.505 & editorial   & 0.549 & concerted      & 0.365 \\
examines     & 0.495 & journalism  & 0.548 & cooperation    & 0.350 \\
discourse    & 0.493 & newspapers  & 0.540 & consultation   & 0.343 \\
theoretical  & 0.488 & tabloid     & 0.530 & achievement    & 0.339 \\
implications & 0.485 & articles    & 0.524 & strengthen     & 0.333 \\
\hline
\end{tabular}
\end{table}

\begin{table}[H]
\centering
\caption{The most similar terms to \textit{cognitive dissonance}.}
\begin{tabular}{l c l c l c}
\hline
\multicolumn{2}{c}{Research} & \multicolumn{2}{c}{News} & \multicolumn{2}{c}{Policy} \\
Term & Similarity & Term & Similarity & Term & Similarity \\
\hline
theories     & 0.690 & sort       & 0.646 & conclusions  & 0.331 \\
theory       & 0.679 & kind       & 0.627 & conditions   & 0.322 \\
hypothesis   & 0.622 & sense      & 0.620 & detailed     & 0.318 \\
hypotheses   & 0.546 & explain    & 0.616 & conclusion   & 0.305 \\
reasoning    & 0.536 & understand & 0.614 & advanced     & 0.289 \\
empirical    & 0.518 & something  & 0.607 & comprehensive& 0.279 \\
cognitive    & 0.507 & indeed     & 0.600 & satisfactory & 0.271 \\
assumptions  & 0.483 & notion     & 0.599 & theories     & 0.269 \\
mechanisms   & 0.475 & fact       & 0.599 & certainty    & 0.268 \\
behaviors    & 0.474 & thing      & 0.596 & complete     & 0.264 \\
\hline
\end{tabular}
\end{table}

\begin{table}[H]
\centering
\caption{The most similar terms to \textit{attitude change}.}
\begin{tabular}{l c l c l c}
\hline
\multicolumn{2}{c}{Research} & \multicolumn{2}{c}{News} & \multicolumn{2}{c}{Policy} \\
Term & Similarity & Term & Similarity & Term & Similarity \\
\hline
perception    & 0.580 & change     & 0.683 & importance   & 0.495 \\
behaviors     & 0.577 & definitely & 0.667 & positive     & 0.397 \\
behavior      & 0.572 & need       & 0.661 & awareness    & 0.395 \\
interaction   & 0.568 & think      & 0.648 & stresses     & 0.387 \\
interactions  & 0.567 & really     & 0.628 & constructive & 0.378 \\
attitudes     & 0.560 & want       & 0.615 & message      & 0.378 \\
negative      & 0.549 & understand & 0.615 & dialogue     & 0.373 \\
responses     & 0.528 & something  & 0.600 & understanding& 0.370 \\
perceptions   & 0.527 & thing      & 0.595 & regards      & 0.367 \\
theory        & 0.525 & going      & 0.594 & urgency      & 0.360 \\
\hline
\end{tabular}
\end{table}

\begin{table}[H]
\centering
\caption{The most similar terms to \textit{diffusion theory}.}
\begin{tabular}{l c l c l c}
\hline
\multicolumn{2}{c}{Research} & \multicolumn{2}{c}{News} & \multicolumn{2}{c}{Policy} \\
Term & Similarity & Term & Similarity & Term & Similarity \\
\hline
theory        & 0.626 & concepts    & 0.381 & technological & 0.430 \\
theories      & 0.575 & concept     & 0.360 & interventions & 0.385 \\
concepts      & 0.558 & ideas       & 0.335 & overcome     & 0.365 \\
methodology   & 0.541 & design      & 0.326 & problems     & 0.341 \\
paradigm      & 0.493 & theory      & 0.316 & developments & 0.340 \\
model         & 0.493 & theories    & 0.296 & focused      & 0.330 \\
concept       & 0.492 & basic       & 0.275 & explain      & 0.321 \\
evolution     & 0.468 & upon        & 0.269 & failure      & 0.317 \\
theoretical   & 0.445 & rate        & 0.260 & innovation   & 0.317 \\
diffusion     & 0.438 & developed   & 0.260 & strategies   & 0.317 \\
\hline
\end{tabular}
\end{table}

\begin{table}[ht]
\centering
\caption{The most similar terms to \textit{channel effect}.}
\begin{tabular}{l c l c l c}
\hline
\multicolumn{2}{c}{Research} & \multicolumn{2}{c}{News} & \multicolumn{2}{c}{Policy} \\
Term & Similarity & Term & Similarity & Term & Similarity \\
\hline
impression   & 0.493 & equities   & 0.520 & economy     & 0.463 \\
impressions  & 0.466 & investors  & 0.510 & affected    & 0.445 \\
media        & 0.365 & haven      & 0.431 & stock       & 0.443 \\
supervisor   & 0.348 & portfolio  & 0.427 & confidence  & 0.437 \\
resulting    & 0.324 & assets     & 0.423 & prices      & 0.410 \\
stimuli      & 0.312 & markets    & 0.403 & decline     & 0.407 \\
evaluations  & 0.310 & safe       & 0.377 & affects     & 0.390 \\
effects      & 0.310 & investment & 0.362 & markets     & 0.382 \\
content      & 0.307 & banks      & 0.329 & visibly     & 0.380 \\
management   & 0.305 & stream     & 0.316 & investors   & 0.370 \\
\hline
\end{tabular}
\label{tab:diffusion}
\end{table}

\begin{longtable}{p{0.08\textwidth} p{0.29\textwidth} p{0.29\textwidth} p{0.29\textwidth}}
\caption{The most similar terms to \textit{social networks} by year.}\label{tab:socnet_yearly}\\
\toprule
\textbf{Year} & \textbf{Research} & \textbf{News} & \textbf{Policy} \\
\midrule
\endfirsthead

\toprule
\textbf{Year} & \textbf{Research} & \textbf{News} & \textbf{Policy} \\
\midrule
\endhead

\midrule
\multicolumn{4}{r}{\emph{Continued on next page}}\\
\bottomrule
\endfoot

\bottomrule
\endlastfoot

2000 & \makecell[l]{environments\\ relationships\\ methods}
     & \makecell[l]{relationships\\ lack\\ understanding}
     & \makecell[l]{traumatised\\ counselling\\ crescent} \\
\hline
2001 & \makecell[l]{theory\\ mathematical\\ interaction}
     & \makecell[l]{many\\ relationships\\ among}
     & \makecell[l]{cohesion\\ intergenerational\\ facilitates} \\
\hline
2002 & \makecell[l]{connections\\ important\\ role}
     & \makecell[l]{working\\ provide\\ living}
     & \makecell[l]{territorial\\ cohesion\\ intergenerational} \\
\hline
2003 & \makecell[l]{relationships\\ contexts\\ context}
     & \makecell[l]{community\\ relationships\\ friends}
     & \makecell[l]{implementing\\ involves\\ programme} \\
\hline
2004 & \makecell[l]{relationships\\ relationship\\ importance}
     & \makecell[l]{helps\\ tend\\ help}
     & \makecell[l]{---} \\
\hline
2005 & \makecell[l]{relationships\\ interaction\\ characteristics}
     & \makecell[l]{communities\\ relationships\\ families}
     & \makecell[l]{presence\\ strengthened\\ community} \\
\hline
2006 & \makecell[l]{interaction\\ communication\\ theory}
     & \makecell[l]{internet\\ online\\ users}
     & \makecell[l]{deplete\\ productive\\ base} \\
\hline
2007 & \makecell[l]{relationships\\ interaction\\ communities}
     & \makecell[l]{internet\\ web\\ online}
     & \makecell[l]{knowledge\\ tools\\ skills} \\
\hline

\end{longtable}

\begin{longtable}{p{0.08\textwidth} p{0.29\textwidth} p{0.29\textwidth} p{0.29\textwidth}}
\caption{The most similar terms to \textit{public sphere} by year.}\label{tab:publicsphere_yearly}\\
\toprule
\textbf{Year} & \textbf{Research} & \textbf{News} & \textbf{Policy} \\
\midrule
\endfirsthead

\toprule
\textbf{Year} & \textbf{Research} & \textbf{News} & \textbf{Policy} \\
\midrule
\endhead

\midrule
\multicolumn{4}{r}{\emph{Continued on next page}}\\
\bottomrule
\endfoot

\bottomrule
\endlastfoot

2000 & \makecell[l]{discourse\\ cultural\\ debates}
     & \makecell[l]{become\\ reluctant\\ many}
     & \makecell[l]{womens\\ events\\ transnational} \\
\hline
2001 & \makecell[l]{discourse\\ habermas\\ sphere}
     & \makecell[l]{debate\\ writers\\ thoughts}
     & \makecell[l]{gender\\ racism\\ racial} \\
\hline
2002 & \makecell[l]{discourse\\ sphere\\ examines}
     & \makecell[l]{politics\\ discourse\\ culture}
     & \makecell[l]{xenophobia\\ racism\\ diversities} \\
\hline
2003 & \makecell[l]{discourse\\ concepts\\ theories}
     & \makecell[l]{morality\\ values\\ democracy}
     & \makecell[l]{international\\ tourism\\ investment} \\
\hline
2004 & \makecell[l]{theories\\ theory\\ concepts}
     & \makecell[l]{belief\\ traditions\\ politics}
     & \makecell[l]{stimulating\\ productivity\\ enhance} \\
\hline
2005 & \makecell[l]{discourse\\ sphere\\ discussion}
     & \makecell[l]{politics\\ respect\\ religion}
     & \makecell[l]{democracy\\ governance\\ legitimacy} \\
\hline
2006 & \makecell[l]{discourse\\ theory\\ theories}
     & \makecell[l]{religion\\ belief\\ notion}
     & \makecell[l]{academics\\ participation\\ debate} \\
\hline
2007 & \makecell[l]{discourse\\ theory\\ theories}
     & \makecell[l]{religion\\ belief\\ regard}
     & \makecell[l]{european\\ citizens\\ promoting} \\
\hline

\end{longtable}

\begin{longtable}{p{0.08\textwidth} p{0.29\textwidth} p{0.29\textwidth} p{0.29\textwidth}}
\caption{The most similar terms to \textit{social capital} by year.}\label{tab:socialcapital_yearly}\\
\toprule
\textbf{Year} & \textbf{Research} & \textbf{News} & \textbf{Policy} \\
\midrule
\endfirsthead

\toprule
\textbf{Year} & \textbf{Research} & \textbf{News} & \textbf{Policy} \\
\midrule
\endhead

\midrule
\multicolumn{4}{r}{\emph{Continued on next page}}\\
\bottomrule
\endfoot

\bottomrule
\endlastfoot

2000 & \makecell[l]{dimensions\\ civic\\ personality}
     & \makecell[l]{local\\ community\\ grants}
     & \makecell[l]{indirectly\\ directly\\ irrespective} \\
\hline
2001 & \makecell[l]{related\\ uses\\ patterns}
     & \makecell[l]{community\\ fund\\ local}
     & \makecell[l]{local\\ organisations\\ human} \\
\hline
2002 & \makecell[l]{erodes\\ eroded\\ questioning}
     & \makecell[l]{communities\\ community\\ trust}
     & \makecell[l]{human\\ governance\\ development} \\
\hline
2003 & \makecell[l]{physical\\ theory\\ proximity}
     & \makecell[l]{communities\\ community\\ prosperity}
     & \makecell[l]{importance\\ improving\\ strengthening} \\
\hline
2004 & \makecell[l]{health\\ social\\ wellbeing}
     & \makecell[l]{communities\\ community\\ social}
     & \makecell[l]{opportunities\\ women\\ equal} \\
\hline
2005 & \makecell[l]{ties\\ relations\\ links}
     & \makecell[l]{communities\\ community\\ importance}
     & \makecell[l]{significant\\ impacts\\ impact} \\
\hline
2006 & \makecell[l]{interaction\\ interactions\\ relationships}
     & \makecell[l]{community\\ communities\\ importance}
     & \makecell[l]{detrimental\\ impact\\ harmful} \\
\hline
2007 & \makecell[l]{interpersonal\\ relationships\\ communication}
     & \makecell[l]{communities\\ community\\ diversity}
     & \makecell[l]{outcomes\\ human\\ ict} \\
\hline

\end{longtable}

\begin{longtable}{p{0.08\textwidth} p{0.29\textwidth} p{0.29\textwidth} p{0.29\textwidth}}
\caption{The most similar terms to \textit{social identity} by year.}\label{tab:socialidentity_yearly}\\
\toprule
\textbf{Year} & \textbf{Research} & \textbf{News} & \textbf{Policy} \\
\midrule
\endfirsthead

\toprule
\textbf{Year} & \textbf{Research} & \textbf{News} & \textbf{Policy} \\
\midrule
\endhead

\midrule
\multicolumn{4}{r}{\emph{Continued on next page}}\\
\bottomrule
\endfoot

\bottomrule
\endlastfoot

2000 & \makecell[l]{theory\\ context\\ studies}
     & \makecell[l]{cultures\\ distinct\\ relationships}
     & \makecell[l]{physical\\ mental\\ processing} \\
\hline
2001 & \makecell[l]{linguistic\\ theory\\ distinct}
     & \makecell[l]{portraits\\ portrait\\ photographs}
     & \makecell[l]{---} \\
\hline
2002 & \makecell[l]{storytelling\\ theory\\ stereotyping}
     & \makecell[l]{feelings\\ relationship\\ sense}
     & \makecell[l]{physical\\ mental\\ physiological} \\
\hline
2003 & \makecell[l]{theory\\ posits\\ postulates}
     & \makecell[l]{british\\ constructing\\ islamic}
     & \makecell[l]{---} \\
\hline
2004 & \makecell[l]{theories\\ theory\\ differences}
     & \makecell[l]{identity\\ distinct\\ unique}
     & \makecell[l]{---} \\
\hline
2005 & \makecell[l]{hypothesis\\ theory\\ theories}
     & \makecell[l]{sexual\\ sense\\ rude}
     & \makecell[l]{physical\\ mental\\ physiological} \\
\hline
2006 & \makecell[l]{theory\\ theorizing\\ communicative}
     & \makecell[l]{identity\\ values\\ culture}
     & \makecell[l]{physical\\ mental\\ physiological} \\
\hline
2007 & \makecell[l]{interactions\\ theory\\ contexts}
     & \makecell[l]{friendship\\ peoples\\ deepen}
     & \makecell[l]{physical\\ emotional\\ cultural} \\
\hline

\end{longtable}
\begin{longtable}{p{0.08\textwidth} p{0.29\textwidth} p{0.29\textwidth} p{0.29\textwidth}}
\caption{The most similar terms to \textit{persuasion} by year.}\label{tab:persuasion_yearly}\\
\toprule
\textbf{Year} & \textbf{Research} & \textbf{News} & \textbf{Policy} \\
\midrule
\endfirsthead

\toprule
\textbf{Year} & \textbf{Research} & \textbf{News} & \textbf{Policy} \\
\midrule
\endhead

\midrule
\multicolumn{4}{r}{\emph{Continued on next page}}\\
\bottomrule
\endfoot

\bottomrule
\endlastfoot

2000 & \makecell[l]{acceptance\\ criticism\\ critical}
     & \makecell[l]{willing\\ whatever\\ take}
     & \makecell[l]{---} \\
\hline
2001 & \makecell[l]{theory\\ cognitive\\ processes}
     & \makecell[l]{give\\ kind\\ whatever}
     & \makecell[l]{odour\\ scent\\ clout} \\
\hline
2002 & \makecell[l]{cognition\\ discourse\\ theory}
     & \makecell[l]{whatever\\ willing\\ give}
     & \makecell[l]{instruments\\ saving\\ devices} \\
\hline
2003 & \makecell[l]{theories\\ theory\\ context}
     & \makecell[l]{whatever\\ give\\ willing}
     & \makecell[l]{willingness\\ flexibility\\ pragmatism} \\
\hline
2004 & \makecell[l]{rhetorical\\ narrative\\ reasoning}
     & \makecell[l]{kind\\ whatever\\ sort}
     & \makecell[l]{greater\\ give\\ reason} \\
\hline
2005 & \makecell[l]{concepts\\ discourse\\ theory}
     & \makecell[l]{rather\\ whatever\\ take}
     & \makecell[l]{necessary\\ intended\\ voluntary} \\
\hline
2006 & \makecell[l]{theory\\ processes\\ context}
     & \makecell[l]{kind\\ little\\ sort}
     & \makecell[l]{must\\ decide\\ exert} \\
\hline
2007 & \makecell[l]{ethical\\ discourse\\ theory}
     & \makecell[l]{rather\\ give\\ little}
     & \makecell[l]{intended\\ needed\\ effort} \\
\hline

\end{longtable}

\begin{longtable}{p{0.08\textwidth} p{0.29\textwidth} p{0.29\textwidth} p{0.29\textwidth}}
\caption{The most similar terms to \textit{minimal effects} by year.}\label{tab:minimaleffects_yearly}\\
\toprule
\textbf{Year} & \textbf{Research} & \textbf{News} & \textbf{Policy} \\
\midrule
\endfirsthead

\toprule
\textbf{Year} & \textbf{Research} & \textbf{News} & \textbf{Policy} \\
\midrule
\endhead

\midrule
\multicolumn{4}{r}{\emph{Continued on next page}}\\
\bottomrule
\endfoot

\bottomrule
\endlastfoot

2000 & \makecell[l]{---}
     & \makecell[l]{increase\\ significantly\\ although}
     & \makecell[l]{gases\\ dry\\ chemical} \\
\hline
2001 & \makecell[l]{---}
     & \makecell[l]{increase\\ drop\\ analysts}
     & \makecell[l]{parallel\\ trade\\ demonstrate} \\
\hline
2002 & \makecell[l]{---}
     & \makecell[l]{strike\\ plans\\ affected}
     & \makecell[l]{gases\\ chemical\\ dryers} \\
\hline
2003 & \makecell[l]{---}
     & \makecell[l]{although\\ still\\ far}
     & \makecell[l]{drying\\ gases\\ considered} \\
\hline
2004 & \makecell[l]{---}
     & \makecell[l]{increase\\ recent\\ expected}
     & \makecell[l]{drying\\ gases\\ dryers} \\
\hline
2005 & \makecell[l]{opinion\\ surveys\\ prevailing}
     & \makecell[l]{result\\ traffic\\ construction}
     & \makecell[l]{drying\\ dryers\\ gases} \\
\hline
2006 & \makecell[l]{---}
     & \makecell[l]{reduce\\ proposed\\ traffic}
     & \makecell[l]{drying\\ dryers\\ gases} \\
\hline
2007 & \makecell[l]{errors\\ script\\ differences}
     & \makecell[l]{plans\\ already\\ supply}
     & \makecell[l]{drying\\ veterinarians\\ farmer} \\
\hline

\end{longtable}

\begin{longtable}{p{0.08\textwidth} p{0.29\textwidth} p{0.29\textwidth} p{0.29\textwidth}}
\caption{The most similar terms to \textit{priming} by year.}\label{tab:priming_yearly}\\
\toprule
\textbf{Year} & \textbf{Research} & \textbf{News} & \textbf{Policy} \\
\midrule
\endfirsthead

\toprule
\textbf{Year} & \textbf{Research} & \textbf{News} & \textbf{Policy} \\
\midrule
\endhead

\midrule
\multicolumn{4}{r}{\emph{Continued on next page}}\\
\bottomrule
\endfoot

\bottomrule
\endlastfoot

2000 & \makecell[l]{framing\\ setting\\ approaches}
     & \makecell[l]{pump\\ needed\\ money}
     & \makecell[l]{---} \\
\hline
2001 & \makecell[l]{semantic\\ syntactic\\ lexical}
     & \makecell[l]{pump\\ money\\ pumps}
     & \makecell[l]{---} \\
\hline
2002 & \makecell[l]{negative\\ subliminal\\ effects}
     & \makecell[l]{pump\\ pumps\\ money}
     & \makecell[l]{---} \\
\hline
2003 & \makecell[l]{theories\\ theory\\ hypothesis}
     & \makecell[l]{pump\\ money\\ pumps}
     & \makecell[l]{promoted\\ partnership\\ engineering} \\
\hline
2004 & \makecell[l]{grammatical\\ lexical\\ semantic}
     & \makecell[l]{pump\\ money\\ needed}
     & \makecell[l]{---} \\
\hline
2005 & \makecell[l]{effects\\ manipulate\\ analysis}
     & \makecell[l]{pump\\ pumping\\ money}
     & \makecell[l]{---} \\
\hline
2006 & \makecell[l]{syntactic\\ lexical\\ linguistic}
     & \makecell[l]{pump\\ money\\ needed}
     & \makecell[l]{---} \\
\hline
2007 & \makecell[l]{framing\\ semantic\\ syntactic}
     & \makecell[l]{pump\\ pumps\\ money}
     & \makecell[l]{---} \\
\hline

\end{longtable}
\begin{longtable}{p{0.08\textwidth} p{0.29\textwidth} p{0.29\textwidth} p{0.29\textwidth}}
\caption{The most similar terms to \textit{agenda setting} by year.}\label{tab:agendasetting_yearly}\\
\toprule
\textbf{Year} & \textbf{Research} & \textbf{News} & \textbf{Policy} \\
\midrule
\endfirsthead

\toprule
\textbf{Year} & \textbf{Research} & \textbf{News} & \textbf{Policy} \\
\midrule
\endhead

\midrule
\multicolumn{4}{r}{\emph{Continued on next page}}\\
\bottomrule
\endfoot

\bottomrule
\endlastfoot

2000 & \makecell[l]{framing\\ theoretical\\ theory}
     & \makecell[l]{newspaper\\ bbc\\ programme}
     & \makecell[l]{research\\ stimulate\\ efforts} \\
\hline
2001 & \makecell[l]{media\\ newscasts\\ primary}
     & \makecell[l]{become\\ debate\\ perhaps}
     & \makecell[l]{---} \\
\hline
2002 & \makecell[l]{issue\\ discussed\\ discourse}
     & \makecell[l]{presenting\\ today\\ news}
     & \makecell[l]{---} \\
\hline
2003 & \makecell[l]{influence\\ impact\\ negative}
     & \makecell[l]{debate\\ news\\ weekly}
     & \makecell[l]{---} \\
\hline
2004 & \makecell[l]{influence\\ politics\\ negative}
     & \makecell[l]{newscasts\\ news\\ radio}
     & \makecell[l]{---} \\
\hline
2005 & \makecell[l]{theories\\ political\\ theory}
     & \makecell[l]{newspaper\\ weekly\\ newspapers}
     & \makecell[l]{---} \\
\hline
2006 & \makecell[l]{theory\\ analysis\\ theories}
     & \makecell[l]{news\\ press\\ journalism}
     & \makecell[l]{---} \\
\hline
2007 & \makecell[l]{framing\\ priming\\ theories}
     & \makecell[l]{televised\\ documentaries\\ bbc}
     & \makecell[l]{regional\\ roadmaps\\ agendas} \\
\hline

\end{longtable}


\begin{table}[H]
\centering
\caption{The most similar terms to \textit{social network} using domain-specific embeddings.}
\begin{tabular}{l c l c l c}
\hline
\multicolumn{2}{c}{Research} & \multicolumn{2}{c}{News} & \multicolumn{2}{c}{Policy} \\
Term & Similarity & Term & Similarity & Term & Similarity \\
\hline
interaction   & 0.747 & facebook   & 0.920 & online    & 0.796 \\
mediated      & 0.703 & networks   & 0.912 & internet  & 0.771 \\
communication & 0.696 & users      & 0.877 & networks  & 0.769 \\
study         & 0.672 & online     & 0.875 & platforms & 0.766 \\
theory        & 0.663 & twitter    & 0.870 & websites  & 0.766 \\
social        & 0.663 & sites      & 0.864 & via       & 0.757 \\
media         & 0.661 & networking & 0.839 & users     & 0.753 \\
implications  & 0.650 & internet   & 0.838 & user      & 0.731 \\
networks      & 0.646 & social     & 0.829 & web       & 0.717 \\
discourse     & 0.639 & media      & 0.812 & network   & 0.713 \\
\hline
\end{tabular}
\label{tab:domain}
\end{table}

\begin{table}[H]
\centering
\caption{The most similar terms to \textit{public sphere} using domain-specific embeddings.}
\begin{tabular}{l c l c l c}
\hline
\multicolumn{2}{c}{Research} & \multicolumn{2}{c}{News} & \multicolumn{2}{c}{Policy} \\
Term & Similarity & Term & Similarity & Term & Similarity \\
\hline
sphere       & 0.788 & sphere      & 0.894 & society        & 0.739 \\
discourse    & 0.763 & politics    & 0.816 & citizens       & 0.732 \\
theory       & 0.631 & political   & 0.799 & participation  & 0.707 \\
examines     & 0.581 & belief      & 0.782 & cultural       & 0.699 \\
explores     & 0.577 & religion    & 0.776 & organisations  & 0.682 \\
mediated     & 0.575 & faith       & 0.774 & european       & 0.677 \\
implications & 0.575 & moral       & 0.772 & sphere         & 0.674 \\
theories     & 0.566 & religious   & 0.771 & institutions   & 0.668 \\
theoretical  & 0.558 & society     & 0.761 & importance     & 0.663 \\
debates      & 0.541 & argue       & 0.759 & civil          & 0.661 \\
\hline
\end{tabular}
\end{table}

\begin{table}[H]
\centering
\caption{The most similar terms to \textit{social capital} using domain-specific embeddings.}
\begin{tabular}{l c l c l c}
\hline
\multicolumn{2}{c}{Research} & \multicolumn{2}{c}{News} & \multicolumn{2}{c}{Policy} \\
Term & Similarity & Term & Similarity & Term & Similarity \\
\hline
mediated       & 0.659 & communities & 0.815 & opportunities & 0.672 \\
interaction    & 0.645 & trust       & 0.773 & participation & 0.671 \\
theory         & 0.637 & capital     & 0.756 & employment    & 0.667 \\
implications   & 0.612 & community   & 0.754 & promoting     & 0.667 \\
relationships  & 0.608 & build       & 0.744 & improving     & 0.663 \\
perceptions    & 0.607 & society     & 0.743 & development   & 0.662 \\
study          & 0.607 & vital       & 0.741 & contribute    & 0.657 \\
findings       & 0.590 & helping     & 0.732 & inclusion     & 0.632 \\
attitudes      & 0.585 & wider       & 0.727 & promote       & 0.629 \\
communication  & 0.579 & helps       & 0.722 & importance    & 0.626 \\
\hline
\end{tabular}
\end{table}

\begin{table}[H]
\centering
\caption{The most similar terms to \textit{social identity} using domain-specific embeddings.}
\begin{tabular}{l c l c l c}
\hline
\multicolumn{2}{c}{Research} & \multicolumn{2}{c}{News} & \multicolumn{2}{c}{Policy} \\
Term & Similarity & Term & Similarity & Term & Similarity \\
\hline
theory       & 0.770 & identity       & 0.635 & physiological & 0.771 \\
mediated     & 0.710 & cultural       & 0.555 & physical      & 0.555 \\
theoretical  & 0.662 & understanding  & 0.534 & processing    & 0.503 \\
attitudes    & 0.626 & religious      & 0.526 & genetic       & 0.430 \\
discourse    & 0.622 & sense          & 0.525 & psychological & 0.400 \\
implications & 0.616 & belonging      & 0.522 & mental        & 0.391 \\
perceptions  & 0.612 & self           & 0.516 & cultural      & 0.367 \\
examined     & 0.608 & relationships  & 0.513 & factors       & 0.363 \\
theories     & 0.606 & culture        & 0.511 & identity      & 0.344 \\
framing      & 0.600 & ones           & 0.508 & natural       & 0.336 \\
\hline
\end{tabular}
\end{table}

\begin{table}[H]
\centering
\caption{The most similar terms to \textit{persuasion} using domain-specific embeddings.}
\begin{tabular}{l c l c l c}
\hline
\multicolumn{2}{c}{Research} & \multicolumn{2}{c}{News} & \multicolumn{2}{c}{Policy} \\
Term & Similarity & Term & Similarity & Term & Similarity \\
\hline
theory       & 0.748 & persuasion & 0.950 & penalties     & 0.635 \\
persuasive   & 0.679 & always     & 0.885 & circumstances & 0.605 \\
narrative    & 0.661 & think      & 0.864 & determine     & 0.592 \\
framing      & 0.644 & certainly  & 0.860 & necessary     & 0.551 \\
mediated     & 0.644 & hard       & 0.857 & appropriate   & 0.543 \\
theoretical  & 0.643 & say        & 0.850 & deterrence    & 0.539 \\
implications & 0.625 & never      & 0.849 & comply        & 0.535 \\
attitudes    & 0.623 & whatever   & 0.847 & case          & 0.533 \\
toward       & 0.618 & enough     & 0.846 & sufficient    & 0.530 \\
behavior     & 0.613 & perhaps    & 0.845 & therefore     & 0.521 \\
\hline
\end{tabular}
\end{table}

\begin{table}[H]
\centering
\caption{The most similar terms to \textit{minimal effect} using domain-specific embeddings.}
\begin{tabular}{l c l c l c}
\hline
\multicolumn{2}{c}{Research} & \multicolumn{2}{c}{News} & \multicolumn{2}{c}{Policy} \\
Term & Similarity & Term & Similarity & Term & Similarity \\
\hline
suggest      & 0.401 & minimal     & 0.809 & drying        & 0.602 \\
findings     & 0.396 & cut         & 0.722 & dryers        & 0.527 \\
study        & 0.394 & cost        & 0.705 & gases         & 0.502 \\
motivations  & 0.389 & increase    & 0.690 & concentration & 0.369 \\
results      & 0.387 & extra       & 0.690 & minimal       & 0.347 \\
activity     & 0.386 & run         & 0.687 & measured      & 0.345 \\
participants & 0.385 & per         & 0.684 & chemical      & 0.344 \\
differences  & 0.382 & high        & 0.683 & acceptable    & 0.300 \\
cognitive    & 0.381 & spending    & 0.682 & gas           & 0.288 \\
attitudes    & 0.377 & plans       & 0.679 & composition   & 0.285 \\
\hline
\end{tabular}
\end{table}

\begin{table}[H]
\centering
\caption{The most similar terms to \textit{social learning} using domain-specific embeddings.}
\begin{tabular}{l c l c l c}
\hline
\multicolumn{2}{c}{Research} & \multicolumn{2}{c}{News} & \multicolumn{2}{c}{Policy} \\
Term & Similarity & Term & Similarity & Term & Similarity \\
\hline
theory        & 0.772 & learning       & 0.747 & competences     & 0.707 \\
theoretical   & 0.664 & teaching       & 0.692 & competence      & 0.693 \\
implications  & 0.634 & students       & 0.666 & learning        & 0.476 \\
theories      & 0.623 & university     & 0.629 & languages       & 0.474 \\
interaction   & 0.587 & spaces         & 0.628 & literacy        & 0.465 \\
framing       & 0.570 & courses        & 0.614 & skills          & 0.445 \\
discourse     & 0.569 & science        & 0.602 & lifelong        & 0.421 \\
attitudes     & 0.566 & learn          & 0.593 & creativity      & 0.411 \\
communication & 0.562 & student        & 0.586 & learners        & 0.391 \\
cognitive     & 0.561 & college        & 0.582 & entrepreneurship & 0.390 \\
\hline
\end{tabular}
\end{table}

\begin{table}[H]
\centering
\caption{The most similar terms to \textit{priming} using domain-specific embeddings.}
\begin{tabular}{l c l c l c}
\hline
\multicolumn{2}{c}{Research} & \multicolumn{2}{c}{News} & \multicolumn{2}{c}{Policy} \\
Term & Similarity & Term & Similarity & Term & Similarity \\
\hline
priming     & 0.615 & pump     & 0.758 & validated    & 0.504 \\
experiment  & 0.608 & priming  & 0.749 & partnership  & 0.408 \\
effects     & 0.562 & boost    & 0.608 & pump         & 0.353 \\
lexical     & 0.549 & start    & 0.576 & committee    & 0.346 \\
theory      & 0.541 & cut      & 0.571 & fund         & 0.343 \\
examined    & 0.512 & money    & 0.560 & engineering  & 0.317 \\
experiments & 0.511 & help     & 0.539 & funds        & 0.298 \\
cognitive   & 0.499 & extra    & 0.533 & partnerships & 0.297 \\
language    & 0.498 & cash     & 0.531 & funding      & 0.292 \\
attitudes   & 0.497 & make     & 0.527 & finalised    & 0.279 \\
\hline
\end{tabular}
\end{table}

\begin{table}[H]
\centering
\caption{The most similar terms to \textit{agenda setting} using domain-specific embeddings.}
\begin{tabular}{l c l c l c}
\hline
\multicolumn{2}{c}{Research} & \multicolumn{2}{c}{News} & \multicolumn{2}{c}{Policy} \\
Term & Similarity & Term & Similarity & Term & Similarity \\
\hline
framing       & 0.716 & newspaper   & 0.765 & fp7           & 0.651 \\
theory        & 0.695 & guardian    & 0.750 & stakeholders  & 0.650 \\
theoretical   & 0.622 & news        & 0.738 & collaboration & 0.619 \\
implications  & 0.619 & agenda      & 0.729 & initiatives   & ---   \\
discourse     & 0.605 & editor      & 0.723 & strategic     & 0.598 \\
theories      & 0.604 & sunday      & 0.718 & innovation    & 0.587 \\
examines      & 0.597 & journalism  & 0.716 & joint         & 0.578 \\
agenda        & 0.591 & press       & 0.705 & partnerships  & 0.576 \\
examined      & 0.586 & newspapers  & 0.703 & cooperation   & 0.576 \\
influence     & 0.576 & readers     & 0.688 & projects      & 0.561 \\
\hline
\end{tabular}
\end{table}

\end{document}